\documentclass[11pt]{article}
\usepackage{amsfonts,colordvi}
\usepackage{amssymb}
\newcommand{\eq}{\begin{equation}}
\newcommand{\en}{\end{equation}}
\newcommand{\eqn}{\begin{eqnarray}}
\newcommand{\enn}{\end{eqnarray}}
\newcommand{\nn}{\nonumber }
\newcommand{\beq}{\begin{equation}}
\newcommand{\eeq}{\end{equation}}
\newcommand{\M}{\ensuremath{\mathcal{M}}}
\newcommand{\CN}{\ensuremath{\mathcal{N}}}
\newcommand{\tM}{\ensuremath{\tilde{M}}}

\newcommand{\tn}{\ensuremath{\tilde{n}}}
\newcommand{\ta}{\ensuremath{\tilde{a}}}
\newcommand{\tb}{\ensuremath{\tilde{b}}}

\newcommand{\tx}{\ensuremath{\tilde{x}}}
\newcommand{\ty}{\ensuremath{\tilde{y}}}
\newcommand{\ti}{\ensuremath{\tilde{I}}}
\newcommand{\tj}{\ensuremath{\tilde{J}}}
\newcommand{\tk}{\ensuremath{\tilde{K}}}
\newcommand{\tl}{\ensuremath{\tilde{L}}}

\newcommand{\tE}{\ensuremath{\tilde{E}}}
\newcommand{\tF}{\ensuremath{\tilde{F}}}

\newcommand{\JTP}[4][{}]{\{ #2\,#3\,#4 \}^{#1}}

\newcommand{\hmu}{\ensuremath{\hat{\mu}}}

\newcommand{\hA}{\ensuremath{{\hat{A}}}}

\newcommand{\tI}{\ensuremath{\tilde{I}}}
\newcommand{\tJ}{\ensuremath{\tilde{J}}}
\newcommand{\tK}{\ensuremath{\tilde{K}}}
\newcommand{\fg}{\ensuremath{\mathfrak{g}}}
\newcommand  {\Rbar} {{\mbox{\rm$\mbox{I}\!\mbox{R}$}}}
\newcommand  {\Hbar} {{\mbox{\rm$\mbox{I}\!\mbox{H}$}}}
\newcommand {\Cbar}{\mathord{\setlength{\unitlength}{1em}
     \begin{picture}(0.6,0.7)(-0.1,0) \put(-0.1,0){\rm C}
        \thicklines \put(0.2,0.05){\line(0,1){0.55}}\end {picture}}}
\begin{document}

\begin{titlepage}

\begin{center}

\hfill SU-ITP-05/25 \vskip 1cm
\begin{LARGE}
\textbf{Unified 
Maxwell-Einstein and Yang-Mills-Einstein \\ \vspace{2mm}
 Supergravity 
Theories
in    Four Dimensions    } \footnote{ Work supported
  in part by the National
Science Foundation under Grant Number PHY-0245337  and by the German
Research Foundation (DFG) under Grant Number ZA 279/1-1.}
\end{LARGE}\\
\vspace{1.0cm}
\begin{large}
Murat G\"{u}naydin$^{\dagger}$ \footnote{murat@phys.psu.edu},
Sean McReynolds$^{\dagger}$ \footnote{sean@phys.psu.edu} and
    Marco Zagermann$^{\ddagger}$     \end{large}\footnote{ zagermann@itp.stanford.edu   }  \\
\vspace{.35cm}
$^{\dagger}$ \emph{Physics Department \\
Pennsylvania State University\\
University Park, PA 16802, USA} \\
\vspace{.3cm}
and \\
\vspace{.3cm} $^{\ddagger}$ \emph{Department of Physics, Stanford University \\
Varian Building, Stanford, CA 94305-4060, USA.} \\
\vspace{0.5cm} {\bf Abstract}
\end{center}
\enlargethispage*{3cm}
\begin{small}

We study   unified $\mathcal{N}=2$ Maxwell-Einstein
supergravity theories  (MESGTs) and unified Yang-Mills Einstein
supergravity theories   (YMESGTs) in four dimensions.  As their  defining property,  these theories
 admit the action of a global
or local     symmetry group
that  is (i) \emph{simple}, and (ii) acts \emph{irreducibly} on  \emph{all} the vector fields
of the theory, including the ``graviphoton''.
 Restricting ourselves to the theories  that originate from  five  dimensions
            via dimensional reduction,
 we find  that the generic Jordan family of MESGTs with the scalar manifolds
  $\mathcal{M}=[SU(1,1)/U(1)] \times [SO(2,n)/SO(2)\times SO(n)]$ are
  all     unified  in four dimensions
  with the unifying global symmetry group $SO(2,n)$.
  Of these theories only one   can be gauged
so as to obtain a unified YMESGT with the gauge group $SO(2,1)$.
Three of the four magical supergravity theories defined by simple
Euclidean Jordan algebras $J_3^{\mathbb{A}} ( \mathbb{A}=
\mathbb{C}, \mathbb{H}, \mathbb{O} )$ of degree 3 are unified
 MESGTs  in  four dimensions.  The   MESGTs  defined by $J_3^{\mathbb{C}}$ and
$J_3^{\mathbb{O}}$ can furthermore be gauged so as to obtain a 4D
unified YMESGT with gauge groups $SO(3,2)$ and   $SO(6,2)$,
respectively. The generic non-Jordan family and the theories whose
scalar manifolds are homogeneous but not symmetric do not lead to
unified  MESGTs in four dimensions. The three infinite families of
unified five-dimensional MESGTs defined by simple Lorentzian Jordan
algebras of degree $p\neq 4$, whose scalar manifolds are
non-homogeneous, do not lead directly to unified MESGTs in four
dimensions under dimensional reduction. However, since their
manifolds are non-homogeneous we are not able to completely rule out
the existence of symplectic sections in which these theories become
unified in four dimensions.

\end{small}

\end{titlepage}

\renewcommand{\theequation}{\arabic{section}.\arabic{equation}}
\section{Introduction}
\setcounter{equation}{0}

As is well-known,      starting from a few well-established
principles, the no-go theorem of Coleman and Mandula  \cite{coma}
  rules out
  the existence of internal ``bosonic''  symmetries
 (continuous  symmetries        described by      Lie groups) that can
interpolate between particles of different spin.
  As is also well-known, supersymmetry circumvents this theorem, because
it is built upon
 ``fermionic''
   symmetry generators with anticommuting parameters and continuous supersymmetry transformations
   that correspond to Lie supergroups
  \cite{wezu,haloso}.

One of the early motivations to   introduce  the concept of
   space-time supersymmetry  into
particle  physics    was    the  hope that it might provide
a link between matter particles and gauge bosons or perhaps
even relate  gravity with the
Yang-Mills interactions of the standard model.
      As we now know,     these ideas cannot be directly realized  in the
  phenomenologically interesting    $\mathcal{N}=1$ supersymmetric  extensions
   of the standard model, because
  \begin{enumerate}
  \item
   no  known standard model particle  in these models
is  related by a
supersymmetry transformation to another standard model particle of      different   spin
\item  the graviton cannot be a
superpartner of a standard model gauge boson in $\mathcal{N}=1$ supergravity.
\end{enumerate}
      In fact,   as
  for unification and simplification, it seems that   the main virtues of
  low energy supersymmetry are the technical resolution of the gauge
  hierarchy problem in grand unified theories   (GUTs)  and
   an improved convergence  of the  three   standard model
couplings at
 the GUT scale  \cite{gutscenario}.

The second point (ii), i.e., the lack of  entanglement (or ``unification'') of
Yang-Mills sectors and gravity in 4D, $\mathcal{N}=1$ supergravity, is  easily traced back to
 the fact that
  both   sectors live in two
different   types of supermultiplets     and that there are no
vector fields in the $\mathcal{N}=1$ supergravity multiplet. This is
no longer true
  in \emph{extended} supergravity
theories, and, in   principle, there could  be
a ``unification'' of Yang-Mills and gravitational interactions
in extended supergravity  due to one or two of the following (well-known)  mechanisms:
\begin{enumerate}
\item Whereas the $\mathcal{N}=1$ supergravity multiplet
contains no physical vector fields, such vector fields (``graviphotons'') do occur in all supergravity multiplets with $\mathcal{N}\geq 2$. These graviphotons can often act as
gauge fields for non-trivial Yang-Mills-type  gauge symmetries,
and if they do, supersymmetry can be viewed as a ``unifying'' symmetry that connects gravity and the corresponding   Yang-Mills interactions.
The prime example of such a theory    in four dimensions     is the $SO(8)$ gauged
$\mathcal{N}=8$ supergravity theory by de Wit and Nicolai \cite{dWN}.
\item For    $\mathcal{N}\leq 4$ supersymmetry, the supergravity multiplet can also be coupled to additional vector multiplets.
In such a situation, supersymmetry, just as before, directly
connects the graviton to the graviphotons,   but,  at the linearized level,
not to the vector fields from the vector multiplets.
 However, the non-linear couplings in all $\mathcal{N}=4$ supergravity theories and in   many  $\mathcal{N}=2$ theories allow
for certain non-compact   global
  symmetries that interpolate between the vector fields
in the supergravity multiplet  and the vector fields of the vector multiplets.
In some cases, these interpolating non-compact symmetries can also be gauged. When this is the case, the noncompact symmetry generators map vector fields from vector multiplets
to vector fields in the supergravity multiplet, which in turn are then linked to the graviton by supersymmetry.

\end{enumerate}

In both of these setups,  a
 particularly symmetric situation  occurs when the Yang-Mills-type gauge group is  (a)
    \emph{simple}    and  (b)
  acts   \emph{irreducibly}
  on (c)  \emph{all}    the vector fields   of the theory.
 For theories of type (i) with only the supergravity multiplet
 present, the above-mentioned $SO(8)$ gauged $\mathcal{N}=8$
 supergravity (as well as its cousins with the non-compact analogues of $SO(8)$)
 form  an example of this highly symmetric type of theories.

For theories of type (ii), in which also vector multiplets are present,  it is much harder to find
examples with \emph{simple} gauge groups that act \emph{irreducibly}
on \emph{all} vector fields (including the ``graviphotons'').
Curiously,  such theories do exist,      however,
  as was first pointed out in \cite{GST2}
   in the context of 5D, $\mathcal{N}=2$ supergravity coupled to vector multiplets.
In \cite{GST2,GZ4}, such supergravity   theories with    vector
multiplets and simple gauge groups that act irreducibly on all
vector fields were called ``unified Yang-Mills-Einstein supergravity
theories'' (unified YMESGTs),  and their existence was studied in
five dimensions \cite{GZ4}. It is easy to see that unified YMESGTs
are impossible  for 5D, $\mathcal{N}=4$ supergravity coupled to
$\mathcal{N}=4$ vector multiplets \cite{5DN4}, leaving
$\mathcal{N}=2$ supergravity theories as the only
 possible
candidates in five dimensions. As was shown in \cite{GZ4}, if one restricts oneself to symmetric or homogeneous target spaces,
there is precisely one
5D,  $\mathcal{N}=2$ unified YMESGT, which  describes the coupling of
14 vector multiplets to supergravity with
 the gauge group $SU(3,1)$ (which is 15-dimensional and, hence also involves the vector field from the gravity multiplet)
  \cite{GST2}.  Dropping this restriction  to  symmetric or homogeneous
 scalar manifolds, however, allows for  an infinite tower of  unified YMESGTs, in which the
  target spaces are non-homogeneous and the gauge groups are of the form $SU(1,N)$ for arbitrary $N\geq 2$.

Turning off the gauge coupling of a  unified YMESGT gives a
``unified Maxwell-Einstein supergravity theory'' (unified MESGT), in
which now a \emph{global} simple symmetry group of the Lagrangian
acts irreducibly on all the vector   fields. By contrast, not all
unified MESGTs can be turned into  unified YMESGTs, as not every
global symmetry group can     be gauged in extended supergravity.
In this sense,    unified YMESGTs   derive,    in general,
    only from
  a subset of the possible  unified MESGTs.
In \cite{GZ4}, it was shown that, in five dimensions,
 there are precisely four unified
MESGTs if one restricts oneself to symmetric or homogeneous target
spaces. These four theories  had been constructed in \cite{GST1}.
Again, relaxing the restriction   to symmetric or homogeneous target
spaces greatly increases the possibilities, and three infinite
families of unified MESGTs as well as one  exceptional one were
found in \cite{GZ4}.    Interestingly,   all the known unified MESGTs and
YMESGTs in five dimensions    are in some way  based on Jordan
algebras, which hints at some deep mathematical structure
that underlies these theories.

 Motivated by the idea that physics may be effectively
five-dimensional    over some range of distance scales, possible compact
gauge groups in  five-dimensional, $\mathcal{N}=2$  supergravity
theories with vector, tensor and hypermultiplets  were studied and
classified in \cite{EGZ}. The unified theories with noncompact gauge
groups constructed in \cite{GZ4} extend greatly the options for
noncompact gauge groups. Some interesting
  applications   of some of the theories constructed and classified  in \cite{EGZ} and  \cite{GZ4} in the
   context of 5D GUT model building along the lines of \cite{fivedguts} have recently been
   studied in \cite{SMcR}     \footnote{See also \cite{DGKL,ZGAZ}}.

Motivated by possible physical applications as well as the goal of
uncovering further their connection to some  deep underlying
mathematical structures, we extend, in this paper, the analysis of
\cite{GZ4}  to $\mathcal{N}=2$ unified MESGTs and unified YMESGTs in
four dimensions. A complete classification of such theories in four
dimensions is complicated by  the fact that
 the 4D $\mathcal{U}$-duality group is in general realized only on-shell and that the prepotentials
in 4D no longer have to be cubic polynomials. In this paper we will
restrict our analysis  to those 4D theories that can be obtained
from  known five-dimensional $\mathcal{N}=2$ MESGTs or YMESGTs by
dimensional reduction.  The theories that have a five-dimensional
origin are physically interesting for a number of reasons. They are
relevant for describing the classical large volume limits  of IIA
string theory compactifications on Calabi-Yau manifolds (or  for
M-theory compactifications  on certain singular Calabi-Yau manifolds
times $S^1$  that lead to       YMESGTs  coupled to hypermultiplets
           \cite{mtheoryoncalabiyau,MZ,LMZ,JMS1,MS1}).    Therefore, these  theories are important for finding
the M-theoretic origins of orbifold-GUTs.
 Furthermore the four-dimensional $\mathcal{N}=2$ no-scale
supergravity models have their origins in five dimensions
\cite{GST5,cremmeretal,noscale}.

The organization of this paper is as follows. In sections 2 and 3 we
review, respectively,  the  $\mathcal{N}=2$   MESGTs and the known classification of unified 5D
MESGTs and YMESGTs. In section 4 we recall  the bosonic
Lagrangian  of the  $\mathcal{N}=2$  MESGTs dimensionally reduced to four
dimensions and discuss the very special geometries of these
theories. Our main results are in section 5 where we give a
classification of the   4D  unified MESGTs and unified YMESGTs that
originate from five dimensions. The conclusion section summarizes
our main results. Appendix A gives a review of basic facts about
Jordan algebras and their symmetries that are relevant to our work.
Appendix B gives a brief review of the symplectic formulation of
four-dimensional $\mathcal{N}=2$  MESGTs.



\section{5D, $\mathcal{N}=2$ Maxwell-Einstein supergravity theories}
\setcounter{equation}{0}

In this section we briefly summarize the structure of general
  5D,  $\mathcal{N}=2$ Maxwell-Einstein
supergravity theories (MESGTs)
\cite{GST1}.\footnote{ Our conventions coincide with those of
ref. \cite{GST1,GST2,GZ1}. In particular, we will use the
mostly positive metric signature $(-++++)$ and impose the
`symplectic' Majorana condition on all fermionic quantities.}

A 5D, $\mathcal{N}=2$ MESGT describes
the coupling of  $\tn$ vector
multiplets to minimal, $\mathcal{N}=2$ supergravity in five dimensions.
The field content  of the  supergravity multiplet  is given by  the
f\"{u}nfbein $e_{\mu}^{m}$, two gravitini $\Psi_{\mu}^{i}$
($i=1,2$) and one vector field $A_{\mu}$ (the graviphoton).
An $\mathcal{N} =2$ vector multiplet contains a
vector field $A_{\mu}$, two spin-$1/2$ fermions $\lambda^{i}$ and
one real scalar field $\varphi$. The fermions of each of these
multiplets transform as doublets under the $USp(2)_{R}\cong
SU(2)_{R}$ R-symmetry group of the $\mathcal{N} =2$ Poincar\'{e}
superalgebra;  all other fields are $SU(2)_{R}$-inert.

Putting everything together, the total field content of an
$\mathcal{N}=2$ MESGT in  5D  is thus
\begin{equation}
\{ e_{\mu}^{m}, \Psi_{\mu}^{i}, A_{\mu}^{\ti}, \lambda^{i\ta}, \varphi^{\tx}\}
\end{equation}
with
\begin{eqnarray*}
\ti&=& 0,1,\ldots, \tn\\
\ta&=& 1,\ldots, \tn\\
\tx&=& 1,\ldots, \tn.
\end{eqnarray*}
As usual,
 we have combined the graviphoton with the $\tn$ vector fields of the $\tn$
vector multiplets into a single $(\tn+1)$-plet of vector fields
$A_{\mu}^{\ti}$ labelled
by the index $\ti$. The indices $\ta, \tb, \ldots$ and $\tx, \ty,
\ldots$ denote the
flat and curved indices, respectively, of the
 $\tn$-dimensional target manifold, $\mathcal{M}$,
of the scalar fields.

The bosonic part of the Lagrangian is
given by (for the fermionic part and further details we refer to   \cite{GST1})
\begin{eqnarray}\label{Lagrange}
e^{-1}\mathcal{L}_{\rm bosonic}&=& -\frac{1}{2}R
-\frac{1}{4}{\stackrel{\circ}{a}}_{\ti\tj}F_{\mu\nu}^{\ti}
F^{\tj\mu\nu}-\frac{1}{2}g_{\tx\ty}(\partial_{\mu}\varphi^{\tx})
(\partial^{\mu}
\varphi^{\ty})+\nonumber \\ &&+
 \frac{e^{-1}}{6\sqrt{6}}C_{\ti\tj\tk}\varepsilon^{\mu\nu\rho\sigma\lambda}
 F_{\mu\nu}^{\ti}F_{\rho\sigma}^{\tj}A_{\lambda}^{\tk},
\end{eqnarray}
where  $e$ and $R$, respectively,
 denote the f\"{u}nfbein determinant and scalar
curvature of spacetime.  $F_{\mu\nu}^{\ti}$ are the
Abelian field strengths of the vector fields $A_{\mu}^{\ti}$. The
metric, $g_{\tx\ty}$, of the scalar manifold $\M$
 and the matrix ${\stackrel{\circ}{a}}_{\ti\tj}$ both depend
on the scalar fields $\varphi^{\tx}$. The completely symmetric
tensor $C_{\ti\tj\tk}$, by contrast, is constant.

As was observed in   reference \cite{GST1}, the entire
$\mathcal{N}=2$ MESGT (including the fermionic terms and the
supersymmetry transformation laws that we have suppressed) is
uniquely determined by    $C_{\ti\tj\tk}$ \cite{GST1}. More
explicitly, $C_{\ti\tj\tk}$ defines  a cubic polynomial,
$\mathcal{V}(h)$,
 in $(\tn+1)$
real variables $h^{\ti}$ $(\ti=0,1,\ldots,\tn)$,
\begin{equation}
\mathcal{V}(h):=C_{\ti\tj\tk}h^{\ti}h^{\tj}h^{\tk}\ .
\end{equation}
This polynomial defines a metric, $a_{\ti\tj}$,
 in the (auxiliary) space $\mathbb{R}^{(\tn+1)}$ spanned by the $h^{\ti}$:
\begin{equation}\label{aij}
a_{\ti\tj}(h):=-\frac{1}{3}\frac{\partial}{\partial h^{\ti}}
\frac{\partial}{\partial h^{\tj}} \ln \mathcal{V}(h)\ .
\end{equation}
The  $\tn$-dimensional    target space, $\mathcal{M}$, of the scalar
fields $\varphi^{\tx}$ can then be represented as the hypersurface
\cite{GST1}
\begin{equation}\label{hyper1}
{\cal V} (h)=C_{\ti\tj\tk}h^{\ti}h^{\tj}h^{\tk}=1 \ ,
\end{equation}
with $g_{\tx\ty}$ being the pull-back of (\ref{aij}) to $\mathcal{M}$:
\begin{equation}
g_{xy}(\varphi)=\frac{3}{2} (\partial_{x} h^{\tilde{I}}) (\partial_{y} h^{\tilde{J}}) a_{\tilde{I} \tilde{J}}|_{\mathcal{V}=1}
\end{equation}
Finally, the quantity ${\stackrel{\circ}{a}}_{\ti\tj}(\varphi)$ appearing in
(\ref{Lagrange}) is given by the componentwise restriction of
$a_{\ti\tj}$ to $\mathcal{M}$:
\[
{\stackrel{\circ}{a}}_{\ti\tj}(\varphi)=a_{\ti\tj}|_{{\cal V}=1} \ .
\]

The physical requirement of the positivity of the kinetic energy
terms for the scalar and vector fields  requires
 $g_{\tx\ty}$ and ${\stackrel{\circ}{a}}_{\ti\tj}$ to  be positive definite.
This requirement induces constraints on the possible $C_{\ti\tj\tk}$,
and in \cite{GST1} it   was shown that any $C_{\ti\tj\tk}$ that
satisfy these constraints can be brought, by a linear transformation,
 to the following form
\begin{equation}\label{canbasis}
C_{000}=1,\quad C_{0ij}=-\frac{1}{2}\delta_{ij},\quad  C_{00i}=0,
\end{equation}
with  the remaining coefficients $C_{ijk}$
 ($i,j,k=1,2,\ldots , \tn$) being  completely arbitrary.
Such a  basis is  called  a ``canonical basis''.
The arbitrariness of the  $C_{ijk}$ in a canonical basis
implies  that,  for a fixed
number $\tn$ of vector multiplets, different
target manifolds $\mathcal{M}$  are, in
general,  possible.


\section{Unified MESGTs in five dimensions}
\setcounter{equation}{0}

In the previous section, we have treated the vector fields from the
vector multiplets and the one that stems from the supergravity
multiplet on an equal footing. \textit{A priori}, this is of course
just a formal procedure that nicely displays the underlying very
special geometry of the theory. As was pointed out already in
\cite{GST1}, however, for some very special tensors $C_{\ti\tj\tk}$,
the corresponding MESGTs  can
have   surprisingly large symmetry groups, $G$, that  are \\
(i)   \emph{simple} and\\
(ii) act \emph{irreducibly}
 on \emph{all} the vector fields, including the ``graviphoton'' \footnote{It should be noted that
 it is the $G$-invariant   linear combination $h_{\tilde{I}}   F_{\mu\nu}^{\tilde{I}}$ that  appears
 in the gravitino transformation law \cite{GST1}. In a given spacetime background the $h_{\tilde{I}}  $
     are constant, in which case  the combination $h_{\tilde{I}} A_{\mu}^{\tilde{I}}$ might  be
called the  ``graviphoton''.
 Although it is not important for this paper, one can think of
  it as the vector field $A_{\mu}^{\ti} c_{\ti}$ where $c_{\ti}$ is the basepoint at which
  ${\stackrel{\circ}{a}}_{\ti\tj}|_c =\delta_{\ti\tj}$ (see  \cite{GST1}).}.\\
As in \cite{GZ4}, we refer to such theories as  ``unified MESGTs''.
More precisely, a unified MESGT is a MESGT in which the coefficients
$C_{\ti\tj\tk}$ form an invariant  third rank tensor of a
\emph{simple} group $G$ that acts \emph{irreducibly} on both the
vector fields $A_{\mu}^{\ti}$ and the embedding functions $h^{\ti}$
of the very special geometry:
\begin{equation} \label{hAtrafo}
h^{\ti}\rightarrow {B^{\ti}}_{\tj}h^{\tj}, \quad A_{\mu}^{\ti}\rightarrow
{B^{\ti}}_{\tj}A_{\mu}^{\tj}
\end{equation}
with
\begin{displaymath}
{B^{\ti'}}_{\ti}{B^{\tj'}}_{\tj}{B^{\tk'}}_{\tk}C_{\ti'\tj'\tk'}
=C_{\ti\tj\tk} \ .
\end{displaymath}
As the entire Lagrangian (\ref{Lagrange})
is uniquely  determined by the $C_{\ti\tj\tk}$, any invariance of the
$C_{\ti\tj\tk}$ is automatically a symmetry of the whole Lagrangian
\footnote{This statement is of course true for all symmetries of the
$C_{\ti\tj\tk}$, independently of whether $G$ is simple or acts irreducibly
on the $A_{\mu}^{\ti}$
and the $h^{\ti}$.} .
On the  scalar manifold $\mathcal{M}$, such  symmetry transformations
 act as isometries,
which becomes evident if one rewrites
the kinetic    term for the scalar fields as \cite{GST1,dWvP1}
\begin{displaymath}
-\frac{1}{2}g_{\tx\ty}(\partial_{\mu}\varphi^{\tx})(\partial^{\mu}
\varphi^{\ty})= \frac{3}{2}C_{\ti\tj\tk}h^{\ti}\partial_{\mu}h^{\tj}
\partial^{\mu}h^{\tk} \ ,
\end{displaymath}
with the $h^{\ti}$ being constrained according to    (\ref{hyper1}).
Hence, finding a unified MESGT is equivalent to
finding an irreducible representation of a simple group $G$
with an invariant symmetric tensor $C_{\ti\tj\tk}$ of rank three
that gives rise to positive definite metrics $g_{xy}$ and
${\stackrel{\circ}{a}}_{\ti\tj}$ (i.e., that  can be brought to the
canonical form (\ref{canbasis})).

Before the work of reference  \cite{GZ4}, only four unified MESGTs
in five dimensions were known. These are  based on scalar manifolds
that are  symmetric spaces :
\begin{eqnarray}
\mathcal{M}&=& SL(3,\mathbb{R})/
SO(3)\qquad
(\tn=5)\nonumber\cr
\mathcal{M}&=& SL(3,\mathbb{C})/
SU(3)\qquad
(\tn=8)\nonumber\cr
\mathcal{M}&=& SU^{*}(6)/
Usp(6)\qquad
(\tn=14)\nonumber\\
\mathcal{M}    &=& E_{6(-26)}/
F_{4}\qquad \qquad
(\tn=26)\label{magicalfamily},
\end{eqnarray}
where we have indicated the number of vector multiplets, $\tn$, for
each of these theories. In these cases, the symmetry groups $G$ of
these theories are simply the isometry groups $SL(3,\mathbb{R})$,
$SL(3,\mathbb{C})$, $SU^{\ast}(6)$ and $E_{6(-26)}$, respectively,
under which  the, respectively, $6$, $9$, $15$ and $27$ vector
fields $A_{\mu}^{\ti}$ transform irreducibly \cite{GST1}. Thus,
according to our definition, all of these four theories are unified
MESGTs.

The fact that these theories are based on symmetric target spaces
certainly makes them a  bit special. However, being a symmetric space and
giving rise to a unified symmetry is by no means the same thing.
For one thing, all the other symmetric spaces that are allowed in
5D, $\mathcal{N}=2$ MESGTs do \emph{not} give rise to unified MESGTs.
Indeed, it is easy to verify that the two
 other existing series of symmetric target spaces \cite{GST1,GST3},
 \eq \mathcal{M} = \
\frac{SO(\tn-1,1)}{SO(\tn-1)}\times SO(1,1), \qquad \tn\geq
1. \label{genericJordan} \en
and
\begin{equation}\mathcal{M}=\frac{SO(1,\tn)}{SO(\tn)},\quad \tn>1 \ ,
\label{genericnonJordan}
\end{equation}
do not give rise to any simple symmetry group that acts irreducibly
on all the vector fields \cite{GST1,GZ4}. Thus, among the theories
based on symmetric spaces, only four are unified MESGTs
(\ref{magicalfamily}).

Moreover,   in  \cite{GZ4}, it was  shown that there can also be
unified MESGTs whose target manifolds are \emph{not} symmetric
spaces. In fact, these form the large majority of unified MESGTs, as
there are infinitely many of them. Interestingly, these novel
unified MESGTs do  have something in common with the theories of the
symmetric families (\ref{magicalfamily}) and (\ref{genericJordan}):
   They   are all related to Jordan algebras, albeit in a
different way than the symmetric space theories
(\ref{magicalfamily}) and (\ref{genericJordan}) \footnote{The third
family of symmetric spaces (\ref{genericnonJordan}), is not related
to Jordan algebras.}.

In order to
   see       this, let us   recall the definition of a Jordan algebra
     (see also  Appendix A):

\vspace{4mm}
\noindent \textbf{Definition 1:}
 A Jordan algebra over a field $\mathbb{F}$ (which we take to be
$\mathbb{R}$ or
$\mathbb{C}$) is an  algebra, $J$, over $\mathbb{F}$
with a symmetric product, $\circ$,
\begin{equation}\label{commute} X\circ Y = Y
\circ X \in J, \quad \forall\,\, X,Y \in J \ ,
\end{equation}
that satisfies
the Jordan identity
\begin{equation}\label{Jidentity}
X\circ (Y \circ X^2)= (X\circ Y) \circ X^2 \ ,
\end{equation}
where $X^2\equiv (X\circ X)$.
\vspace{-1mm}\\

The Jordan identity (\ref{Jidentity}) is automatically satisfied
when the product $\circ$ is associative, but (\ref{Jidentity}) does
not imply associativity. In other words, a Jordan algebra is
commutative, but in general not associative. It is easy to verify
that the Hermitian  operators acting in a Hilbert space form a
Jordan   algebra with the product ``$\circ$'' being one half the
anticommutator \footnote{The prefactor one half in front of the
anticommutator   allows one to represent the identity element by a
unit matrix for finite dimensional Jordan algebras. For the
importance of the prefactor one half for different formulations of
Jordan algebras see the book by McCrimmon \cite{mccrimmon}.}.
   In fact, historically,    Jordan algebras were    introduced
   in an attempt to generalize the formalism of quantum mechanics by
capturing the algebraic essence of Hermitian operators corresponding
to observables without reference to the underlying Hilbert space on
which they act \cite{JvNW}.

A particular subclass of Jordan algebras         is formed by the
  \emph{Euclidean} Jordan algebras:\\
    \vspace{-1mm}\\
\textbf{Definition 2:} A \emph{Euclidean} (or \emph{formally real})
 Jordan algebra is a Jordan algebra for which
the condition $X\circ X + Y\circ Y=0$ implies that $X=Y=0$ for all
$X,Y\in J$. The automorphism groups of Euclidean Jordan algebras
are always compact, which suggests the alternate name ``compact Jordan algebras" as is sometimes used in the literature. \\
\vspace{-1mm}\\
For every Jordan algebra $J$, one can define a norm
form, $N:J\rightarrow \mathbb{R}$, that satisfies the
composition property \cite{jacobson}
\begin{equation}\label{Norm}
N[2X\circ(Y\circ X)-(X\circ X)\circ Y]=N^{2}(X)N(Y).
\end{equation}
The degree, $p$, of the norm form is defined by $N(\lambda X)=\lambda^p N(X)$,
where $\lambda\in \mathbb{R}$; $p$ is also called
 the degree of the Jordan algebra.\\

It was shown in \cite{GST1} that the norm form, $N$, of a Euclidean
Jordan algebra of degree three ($p=3$) can be identified  with the
cubic polynomial, $\mathcal{V}$, of a physically sensible MESGT.    More
  precisely, the   cubic polynomials
of Euclidean Jordan algebras of degree three  satisfy all the constraints
 required by  supersymmetry and the   positivity of kinetic terms.
In addition, these polynomials  satisfy another
    relation,  called the  adjoint identity, which implies that
the corresponding scalar manifolds are symmetric spaces of the form
\[ \mathcal{M}= \frac{\textrm{Str}_0(J)}{\textrm{Aut}(J)} \]
where $\textrm{Str}_0(J)$ and $\textrm{Aut}(J)$ denote, respectively, the reduced structure and
automorphism groups of the underlying Jordan algebra\footnote{The reduced structure group
$\textrm{Str}_{0}(J)$ is simply
the invariance group of the norm form, $N$, of the corresponding
Jordan algebra $J$. As such, it is, for the above Jordan algebras,
 isomorphic to the symmetry group   $G$  of the
corresponding MESGTs.} \cite{GST1,KMC}. The MESGTs
whose cubic polynomials arise in this way, are precisely the first
two families (\ref{magicalfamily}) and (\ref{genericJordan}). The
relevant Jordan algebras are
\begin{enumerate}
 \item The ``magical
Jordan family'' (\ref{magicalfamily}) \footnote{The name ``magical'' derives from the deep
connection with the ``magic square'' of Freudenthal,
Rosenfeld and Tits \cite{FRT}.}
 corresponds to the four \emph{simple} Euclidean Jordan algebras
of degree 3. These simple Jordan algebras are denoted by $J_3^{\mathbb{R}}$,
$J_3^{\mathbb{C}}$, $J_3^{\mathbb{H}}$, $J_3^{\mathbb{O}}$ and
are isomorphic to the Hermitian $(3\times 3)$-matrices over the four
division algebras $\mathbb{R}, \mathbb{C},
\mathbb{H}, \mathbb{O}$ with the product being one half the anticommutator:
\begin{eqnarray}\label{magicals}
J_{3}^{\mathbb{R}}:\quad \mathcal{M}&=& SL(3,\mathbb{R})/
SO(3)\qquad
(\tn=5)\nonumber\\
J_{3}^{\mathbb{C}}:\quad \mathcal{M}&=& SL(3,\mathbb{C})/
SU(3)\qquad
(\tn=8)\nonumber\\
J_{3}^{\mathbb{H}}:\quad \mathcal{M}&=& SU^{*}(6)/ USp(6)\qquad
(\tn=14) \nonumber \\
J_{3}^{\mathbb{O}}:\quad \mathcal{M}&=& E_{6(-26)}/
F_{4}\qquad \qquad
(\tn=26) \ .
\end{eqnarray}
The cubic norm form, $N$, of these Jordan algebras is given by the
determinant  of the corresponding Hermitian $(3\times 3)$-matrices.

\item  For  the ``generic
Jordan family'' (\ref{genericJordan}),   the scalar manifolds factorize:
\[ \mathcal{M} =
\frac{SO(\tilde{n}-1,1)}{SO(\tilde{n}-1)}\times
SO(1,1) \ .
\]

This is also reflected in the underlying Jordan algebras, which are now  reducible and of the form  $J=\mathbb{R}\oplus \Sigma_{\tn}$.
 Here, $\Sigma_{\tn}$ is a Jordan algebra of degree $p=2$
associated with a quadratic norm form in $\tn$ dimensions that has a
``Minkowskian signature'' $(+,-,\ldots,-)$. A simple realization of
$\Sigma_{\tn}$ is provided by $(\tn-1)$ Dirac gamma matrices
$\gamma^i$ , where $ i,j,\ldots=1,\ldots,(\tn-1)$ of an $(\tn-1)$
dimensional Euclidean space together with the identity matrix $
\gamma^0 = \mathbf{1}$,  and the Jordan product $\circ$ being one
half the anticommutator:

\begin{eqnarray}
\gamma^i \circ \gamma^j &=& \frac{1}{2}
\{\gamma^i,\gamma^j\}= \delta^{ij} \mathbf{1}
\nonumber \\
\gamma^0 \circ \gamma^0 &=& \frac{1}{2}
\{\gamma^0,\gamma^0\}= \mathbf{1}\nonumber\\
\gamma^i \circ \gamma^0 &=& \frac{1}{2}
\{\gamma^i,\gamma^0\}= \gamma^i \ .
\end{eqnarray}
  The norm of a general element $X = X_0 \gamma^0 + X_i \gamma^i $ of
$\Sigma_{\tn}$
is defined as
\[  N(X) = \frac{1}{2^{[\tn/2]}} \textrm{Tr}
X \bar{X} = X_0X_0 - X_iX_i \ ,  \] where
\[ \bar{X} \equiv  X_0 \gamma^0 - X_i \gamma^i  \ . \]
 The norm of a general element $y \oplus X $ of the non-simple
Jordan algebra $J=\mathbb{R}\oplus \Sigma_{\tn}$ is simply given by
$y N(X)$ .

\end{enumerate}

Thus, identifying the cubic polynomial $\mathcal{V}$ defined by the
supergravity
coefficients $C_{\ti\tj\tk}$ with the norm form, $N$, of a Euclidean Jordan
algebra  of degree three yields precisely
 the MESGTs based on the scalar manifolds
(\ref{magicalfamily}) and (\ref{genericJordan}).

Let us now describe how the novel unified MESGTs found in \cite{GZ4}
are related to Jordan algebras. In those theories, one does not
relate the $C_{\ti\tj\tk}$ to the \emph{norm form}, $N$, of a
\emph{Euclidean}
 Jordan algebra
of degree three, but instead one relates them to the \emph{structure
constants} of a \emph{Lorentzian} Jordan algebra \footnote{ In
\cite{GZ4} the Lorentzian Jordan algebras were called Minkowskian
Jordan algebras.}

Let us make this more precise.   We first note   that all
the Jordan algebras of $(n\times n)$ Hermitian matrices over
various division algebras are  Euclidean (compact) Jordan
algebras,  which implies that  their  automorphism groups  are compact. In particular, the Jordan
algebras of degree 3 discussed in   (i) and (ii)  are all Euclidean.
The three Jordan algebras  $J_{3}^{\mathbb{A}=\mathbb{R},\mathbb{C},\mathbb{H}  }$ described in (i) are just special cases of the Jordan algebras  of the
  $(n\times n)$ Hermitian  matrices  over $\mathbb{R},\mathbb{C},\mathbb{H}$
  \footnote{ Hermitian
$(n\times n)$ matrices over the octonions do not form Jordan
algebras for $n > 3$.}. These higher-dimensional   analogues are
also all Euclidean Jordan algebras with compact automorphism groups.
Their norm forms, $N$, are again given by the determinants of the
matrices, which implies that the degree of $J_{n}^{\mathbb{A}}$ is
$n$, singling out $J_{3}^{\mathbb{A}}$ as the only  members of these
families with cubic norms.

Now, the  Euclidean Jordan algebras
$J_{n}^{\mathbb{A}=\mathbb{R},\mathbb{C},\mathbb{H}}$  and $J_3^{
\mathbb{O}}$  also have non-Euclidean analogues
$J_{(q,n-q)}^{\mathbb{A}=\mathbb{R},\mathbb{C},\mathbb{H}} $
 (again with $n \geq 3$ ) and  $J_{(1,2)}^{ \mathbb{O}}$, respectively.
 These
non-Euclidean Jordan algebras   are realized by matrices that are
Hermitian with respect to a non-Euclidean ``metric'' $\eta $ with
signature $(q,n-q)$:
\begin{equation}\label{eta}
(\eta X)^\dag = \eta X  \hspace{1cm} \forall  X\in J_{(q,n-q)}^{\mathbb{A}}
\ .
\end{equation}
Obviously, if we choose $\eta$ to have a Euclidean signature,
we obtain back the  Euclidean (compact)
Jordan algebras.

Consider now Lorentzian  Jordan algebras $J_{(1,N)}^{\mathbb{A}} $
with  degree
 $p  =N+1$ defined by choosing $\eta$ to be the
Lorentzian (or   ``Minkowskian'')  metric $\eta = (-,+,+,...,+)$. A general
element, $U$, of $J_{(1,N)}^{\mathbb{A}} $ can be written in the
form

\begin{equation}
U=\left(\begin{array}{cc}
x & -Y^{\dagger}\\
Y & Z
\end{array}\right),
\end{equation}
where $Z$ is an element of the Euclidean subalgebra
$J_{N}^{\mathbb{A}} $ (i.e., it is a Hermitian $(N\times N)$-matrix over
$\mathbb{A}$), $x\in\mathbb{R}$, and $Y$ denotes an
$N$-dimensional  column vector over $\mathbb{A}$.
Under their (non-compact)   automorphism group, $\textrm{Aut}(J_{(1,N)}^{\mathbb{A}})$,
   these  simple Jordan algebras $J_{(1,N)}^{\mathbb{A}}$ decompose into an
irreducible representation formed by the traceless elements
plus a singlet, which is given by the identity element of
$J_{(1,N)}^{\mathbb{A}}$
(i.e., by the unit matrix $U=\mathbf{1}$):
\begin{equation}
J_{(1,N)}^{\mathbb{A}}= \mathbf{1}\oplus \{ \textrm{traceless
elements}\} \equiv  \mathbf{1} \oplus J_{(1,N)_0}^{\mathbb{A}}  .
\end{equation}
The traceless elements do not close under the Jordan product, $\circ$,
but one can define a symmetric product, $\star$, under
 which
the traceless elements do  close as follows: \footnote { Such a product was
introduced among the Hermitian generators
of $SU(N)$ by Michel and Radicati some time ago \cite{mira}.
 Note that Hermitian generators of $SU(N)$ are in
one-to-one correspondence with the traceless elements of
 $J_N^{\mathbb{C}}$.  The ``symmetric'' algebras with the star product $\star$
do not have an identity element.}

\[ A \star B := A \circ B - \frac{1}{(N+1)} \textrm{tr}(A \circ B)
\mathbf{1} \  \] where $ A,B \in  J_{(1,N)_0}^{\mathbb{A}}$ and
$\circ$ is the Jordan product \[ A \circ B = \frac{1}{2} ( AB+BA)\ .
\]
 Thus, the structure constants ($d$-symbols)  of the traceless elements
under the symmetric $ \star $ product will be invariant tensors of
the automorphism groups $\textrm{Aut}(J_{(1,N)}^{\mathbb{A}})$
of the Jordan algebras. Denoting
the traceless elements as
$T_{\ti}$ $(\ti=0,\ldots, (D-2))$ with $D$ being the dimension of
$J_{(1,N)}^{\mathbb{A}} $, we have
\[
T_{\ti} \star T_{\tj} =d_{\ti\tj}^{~~~\tk} T_{\tk} \ .
\]
The $d$-symbols are then given by

\begin{equation}\label{dsymbols}
d_{\ti\tj\tk}\equiv d_{\tj\tk}^{~~~\tl} \tau_{\tl\ti} = \frac{1}{2}
\textrm{tr}(T_{\ti}\{ T_{\tj}, T_{\tk}\})=
\textrm{tr} (T_{\ti}\circ (T_{\tj}\circ
T_{\tk}))
\end{equation}
where \footnote{One can choose the elements $T_{\ti}$ such that
$\tau_{\ti\tj} = \delta_{\ti\tj} $
($\tau_{\ti\tj} = - \delta_{\ti\tj} $)  for two compact (noncompact)
elements $ T_{\ti}$ and $T_{\tj}$ and zero
otherwise.}
\[\tau_{\tl\ti} = \textrm{tr} (T_{\tl}\circ T_{\ti}) \ . \]
The $d_{\ti\tj\tk}$ are completely symmetric in their indices, and as
$\textrm{Aut}(J_{(1,N)}^{\mathbb{A}})$ acts irreducibly on the traceless
elements $T_{\ti}$, the $d_{\ti\tj\tk}$ are a promising candidate
for the $C_{\ti\tj\tk}$ of a unified MESGT. What remains to check, however,
 is
whether the metrics $g_{\tx\ty}$ and
${\stackrel{\circ}{a}}_{\ti\tj}$ on the resulting scalar manifold
$\M$ are really positive definite. It was shown in \cite{GZ4} that
this is true if and only if the signature of $\eta$ is really
$(-,+,\ldots,+)$.

Thus, putting everything together, if one identifies the $d$-symbols
(\ref{dsymbols}) of the traceless elements of a Lorentzian Jordan
algebra $J_{(1,N)}^{\mathbb{A}}$ with the $C_{\ti\tj\tk}$ of a
MESGT: $C_{\ti\tj\tk}=d_{\ti\tj\tk}$, one obtains a unified MESGT,
in which all the vector fields transform irreducibly under the
simple automorphism group $\textrm{Aut}(J_{(1,N)}^{\mathbb{A}})$ of
that Jordan algebra.

For $\mathbb{A}=\mathbb{R},\mathbb{C},\mathbb{H}$ one obtains in
this way three infinite families of physically acceptable unified
MESGTs (one for each $N\geq2$). For the octonionic case, the
situation is a bit different. The $d$-symbols of the octonionic
Lorentzian Hermitian $(N+1)\times (N+1)$-matrices with the
anticommutator product all lead to positive definite metrics
$g_{\tx\ty}$ and ${\stackrel{\circ}{a}}_{\ti\tj}$, i.e., to
physically acceptable MESGTs. For $N\neq  2$, however, these
octonionic Hermitian matrix algebras are no longer Jordan algebras.
 Surprisingly, the automorphism groups of these octonionic  algebras
  for $N\geq 3$ do not contain the automorphism group
 $F_{4(-20)}$ of $J_{(1,2)}^{\mathbb O}$ as a subgroup \cite{goodaire}.
Instead, the  automorphism groups for $N \geq 3$
 are direct product groups of the form  $SO(N,1) \times G_2$.  None
of these two factors acts irreducibly on all the traceless elements, and hence
  the corresponding $\mathcal{N}=2$ MESGTs are  not unified
 theories. Thus, the $\CN=2$ MESGT defined by the exceptional
 Lorentzian
Jordan algebra $J_{(1,2)}^{\mathbb{O}}$ is the only \emph{unified}
MESGT of this infinite tower of otherwise acceptable octonionic theories.

All these results are summarized in Table 1, which  lists all the
simple Lorentzian Jordan algebras of type $J_{(1,N)}^{\mathbb{A}} $,
their  automorphism  groups, and the numbers of vector and scalar
fields in the unified MESGTs defined by them.

 \begin{table}[ht]

\begin{center}
\begin{displaymath}
\begin{array}{|c|c|c|c|c|}
\hline
~&~&~&~&~\\
J & D&  \textrm{Aut}(J)
& \textrm{No. of vector fields} & \textrm{No.  of scalars}  \\
\hline
~&~&~&~&~\\
J_{(1,N)}^{\mathbb R}& \frac{1}{2} (N+1)(N+2) & SO(N,1) &
\frac{1}{2}N(N+3) &\frac{1}{2}N(N+3)-1 \\
~&~&~&~&~\\
J_{(1,N)}^{\mathbb C} & (N+1)^2 & SU(N,1) & N(N+2) & N(N+2)-1 \\
~&~&~&~&~\\
J_{(1,N)}^{\mathbb H} &(N+1)(2N+1) & USp(2N,2) & N(2N+3) & N(2N+3)-1 \\
~&~&~&~&~\\
J_{(1,2)}^{\mathbb O} & 27& F_{4(-20)} & 26 & 25 \\
~&~&~&~&~ \\
\hline
\end{array}
\end{displaymath}
\end{center}

\caption{List of the simple Lorentzian Jordan algebras of type
$J_{(1,N)}^{\mathbb{A}}$.
 The columns show, respectively, their  dimensions $D$,
their automorphism groups $\textrm{Aut}
(J_{(1,N)}^{\mathbb{A}})$,
the number of vector fields $(\tn+1)=(D-1)$
and the number of scalars $\tn=(D-2)$ in the corresponding MESGTs.}
\end{table}

Note that the number of vector fields for the theories defined by
$J_{(1,3)}^{\mathbb R}$, $J_{(1,3)}^{\mathbb C}$ and
$J_{(1,3)}^{\mathbb H}$ are  9, 15 and 27, respectively. These are
exactly the same numbers of vector fields  in the magical theories
based on the norm forms of the Euclidean Jordan algebras
$J_{3}^{\mathbb C}$, $J_{3}^{\mathbb H}$ and $J_{3}^{\mathbb O}$,
respectively. As was shown   in \cite{GZ4}, this is not an accident;
the magical MESGTs based on $J_{3}^{\mathbb C}$, $J_{3}^{\mathbb H}$
and $J_{3}^{\mathbb O}$ found in \cite{GST1} are \emph{equivalent}
(i.e. the cubic polynomials $\mathcal{V}(h)$ agree) to the ones
defined by the Lorentzian algebras $J_{(1,3)}^{\mathbb R}$,
$J_{(1,3)}^{\mathbb C}$ and $J_{(1,3)}^{\mathbb H}$, respectively.
This is a consequence of the fact that the generic norms of  the
degree 3 simple Euclidean Jordan algebras $J_{3}^{\mathbb C}$,
$J_{3}^{\mathbb H}$ and $J_{3}^{\mathbb O}$ coincide with the cubic
norms defined over  the traceless elements of degree four simple
Lorentzian Jordan algebras over $\mathbb{R}, \mathbb{C}$ and
$\mathbb{H}$ \cite{GZ4,alfa}. This implies that the only known
unified MESGT that is not covered by the above table is the magical
theory of \cite{GST1} based on the Euclidean Jordan algebra
$J_{3}^{\mathbb{R}}$ with $(\tn+1)=6$ vector fields and the target
space $\M=SL(3,\mathbb{R})/SO(3)$.

Of these unified MESGTs only the family defined by $
J_{(1,N)}^{\mathbb{C}}$ can be gauged so as to obtain an infinite
family of unified YMESGTs with the gauge groups $SU(N,1)$
\cite{GZ4}.

 Except for the theories defined by $J_{(1,3)}^{\mathbb R}$, $J_{(1,3)}^{\mathbb C}$ and
$J_{(1,3)}^{\mathbb H}$ the scalar manifolds of  MESGTs defined by
other simple Lorentzian Jordan algebras  are not homogeneous.
Let us therefore close this section with a short description of their geometry.

To this end, one recalls   that
the  scalar manifolds $\mathcal{M}$ of Lorentzian Jordan algebraic
 theories    in five dimensions are cubic surfaces defined by the
condition

\[ {\cal V} (h)=C_{\ti\tj\tk}h^{\ti}h^{\tj}h^{\tk}=1,  \]
where $C_{\ti\tj\tk}$ are the structure constants
$C_{\ti\tj\tk}\equiv  d_{\ti\tj\tk}$ of the traceless elements of
the Lorentzian Jordan algebra $J_{(1,N)}^{\mathbb{A}}$.

Let us now show that the scalar manifolds of  5D  unified MESGTs
defined by Lorentzian Jordan algebras $J_{(1,N)}^{\mathbb{A}} $ for
$N \geq 2$ can be embedded in the coset spaces
\[ \mathcal{R} = \frac{\textrm{Str}_0(J_{(1,N)}^{\mathbb{A}})}{\textrm{Aut}(J_{(1,N)}^{\mathbb{A}})}
\]
where, just as above,  $\textrm{Str}_0(J)$ and $\textrm{Aut}(J)$ denote, respectively, the reduced structure and
automorphism groups of the  Jordan algebra $J$. First the coset space
$\mathcal{R}$ itself can be realized as a
   hypersurface in an  $(\tn+2)$-dimensional  ambient
space with coordinates $\xi^A$ $( A,B=0,1,..,\tn +1)$ and metric

\begin{equation}\label{GAB}
G_{AB}(\xi):=-\frac{1}{3}\frac{\partial}{\partial \xi^A}
\frac{\partial}{\partial \xi^B} \ln \mathcal{P}(\xi)\ ,
\end{equation}
where
\[ \mathcal{P}(\xi) = C_{AB\ldots D}\;\xi^A \xi^B\cdots\xi^D \]
is the generic norm form of the $(\tn+2)$-dimensional Jordan algebra
$J_{(1,N)}^{\mathbb{A}}$ defined by a completely symmetric tensor
$C_{AB\ldots D}$ of rank $N+1$.

 The  $(\tn +1)$-dimensional hypersurface, $\mathcal{R}$, is defined
 by the condition
\begin{equation}\label{hyper}
\mathcal{P}=C_{AB\ldots D}\;\xi^A \xi^B\cdots\xi^D=1 \ .
\end{equation}
Since the norm forms of the Jordan algebras $J_{(1,N)}^{\mathbb{A}}$
 are determinental forms, the norm $\mathcal{P}$ of an element $J$
can be expanded in terms of the products of  traces,
$\textrm{Tr}(J^m)$  of the powers of $J$ ($m=1,2,\ldots,N+1$). A
general element, $J$,  of $J_{(1,N)}^{\mathbb{A}}$ can be decomposed
as \[ J= \xi^{\tn+1 } 1 + e_{\ti} h^{\ti}
\] where $1$ is the identity element and $e_{\ti}$ (with $\ti
=0,\ldots \tn+1$) form a basis of traceless elements of
$J_{(1,N)}^{\mathbb{A}}$ \footnote{ In terms of the coordinates
$\xi^A$($A=0,1,...,\tn+1 )$ we are identifying $\xi^{\ti}$ with
$h^{\ti}$ ($ \ti =0,1,...,\tn )$.}. Now the norm $\mathcal{P}$
depends on one more parameter, namely $\xi^{\tn +1}$, than the cubic
form $\mathcal{V}$ and hence one can simultaneously impose the
conditions
\[ \mathcal{P}=1\;\;\;\;\mbox{and}\;\;\;\;\mathcal{V}=1,\]
showing  that the scalar manifold of a unified MESGT   defined by a
Lorentzian Jordan algebra can be mapped into a cubic hypersurface in
the coset space
 \[ \mathcal{R}=\frac{\textrm{Str}_0(J_{(1,N)}^{\mathbb{A}})}{\textrm{Aut}(J_{(1,N)}^{\mathbb{A}})} .
 \]

We list the coset spaces $\mathcal{R}$ defined by Lorentzian Jordan
algebras   below:

\[
 \mathcal{R}(J_{(1,N)}^{\mathbb{R}}) =
\frac{SL(N+1,\mathbb{R})}{SO(N,1)}  \]
\[\mathcal{R}(J_{(1,N)}^{\mathbb{C}}) =
\frac{SL(N+1,\mathbb{C})}{SU(N,1)}  \]
\[\mathcal{R}(J_{(1,N)}^{\mathbb{H}}) =
\frac{SU^*(2N+2)}{USp(2N,2)} .  \]

\section{Dimensional reduction  of 5D, $\mathcal{N}=2$
  MESGTs to   four dimensions}

In the previous section we have listed the known  unified  MESGTs in five dimensions, as given in \cite{GZ4}.
Interestingly, all of these theories are related to Jordan algebras, but only four of them are based on symmetric
 target spaces. It is the purpose of this paper, to study unified MESGTs (and YMESGTs) in four space-time dimensions.
  A complete classification of these theories would be quite a complicated problem  due to two novel features of
   MESGTs in four dimensions: \\
(i) The prepotentials of the special geometry are, in general,  no
longer purely
cubic polynomials. \\
(ii)  They are subject to symplectic reparametrizations that may
lead to  Lagrangians with  different symmetries.\\
In order to cope with the first difficulty, we will restrict
ourselves to those 4D theories that can be obtained from five
dimensions  by dimensional reduction.  Some of the physical
motivations for this were given above. The resulting ``very
special'' K\"{a}hler manifolds in 4D are then based on cubic
prepotentials.

In this section, we will briefly review  the dimensional reduction of the
  5D, $\mathcal{N}=2$  MESGTs to four
dimensions. As we did in the above section we shall restrict ourselves to the purely bosonic sector of these
theories. We should stress again  that by far  not all  four-dimensional MESGTs   can be obtained by dimensional reduction from
five dimensions. The metric of the target space of the four-dimensional scalar fields of dimensionally reduced
theories were first obtained in \cite{GST1} where a complete list of the symmetric
target manifolds of theories defined by
Jordan algebras were also given.  These manifolds as well as those of the generic non-Jordan family \cite{GST3}
 were later
studied in \cite{dWvP1,dWvP2}, where it was pointed out that not all
the isometries of the scalar manifolds of the generic non-Jordan family
are symmetries of the Lagrangian.

  As in \cite{GST1}, we
  choose the parametrization for the f\"{u}nfbein to be \footnote{In
the rest of the paper we shall use hatted indices for the 5D vectors, i.e $\hat{\mu}, \hat{\nu},\ldots= 0,1,2,3,4 $
and use unhatted indices for four-dimensional vectors, i.e $ \mu, \nu, \ldots=0,1,2,3$ .} \eq
\hat{e}^{\hat{m}}_{\hat{\mu}}= \left( \begin{array}{ccc}
e^{-\frac{\sigma}{2} }e^{m}_{\mu} & & 2 e^{\sigma}  W_{\mu} \\
0 & & e^{\sigma}
\end{array} \right), \en
and  decompose
the vector fields $\hA_{\hmu}^{\tI}$  into a 4D vector field, $A_{\mu}^{\tI}$, and a 4D scalar, $A^{\tI}$,
\begin{equation}
\hA_{\hmu}^{\tI}=  \left( \begin{array}{c}
\hA_{\mu}^{\tI}\\
\hA_{5}^{\tI}
\end{array} \right) = \left( \begin{array}{c}
A_{\mu}^{\tI}+2W_{\mu}A^{\tI}\\
A^{\tilde{I}} \end{array}    \right). \label{vector}
\end{equation}

 Using  this             decomposition,
   the dimensionally reduced Lagrangian becomes
\begin{eqnarray}\label{redlag}
e^{-1}\mathcal{L} &  =
&-\frac{1}{2}R-\frac{3}{4}\partial_{\mu}\sigma \partial^{\mu}\sigma
-\frac{1}{2}e^{-2\sigma}\stackrel{\circ}{a}_{\ti\tj}\partial_{\mu}A^{\ti}\partial^{\mu}A^{\tj}
-\frac{1}{2}g_{\tx\ty}\partial_{\mu}\phi^{\tx}\partial^{\mu}\phi^{\ty}
\\ \nonumber  & &-(\frac{1}{2}e^{3\sigma}+
e^{\sigma}\stackrel{\circ}{a}_{\ti\tj}A^{\ti}A^{\tj})W_{\mu\nu}W^{\mu\nu}
-\frac{1}{4} e^{\sigma}
\stackrel{\circ}{a}_{\ti\tj}F^{\ti}_{\mu\nu}F^{\mu\nu\,\tj} -
e^{\sigma}\stackrel{\circ}{a}_{\ti\tj}A^{\ti}F^{\tj}_{\mu\nu}W^{\mu\nu}
\\ \nonumber &
&+\frac{e^{-1}}{2\sqrt{6}}C_{\ti\tj\tk}\epsilon^{\mu\nu\rho\sigma}(A^{\ti}F^{\tj}_{\mu\nu}F^{\tk}_{\rho\sigma}
+2A^{\ti}A^{\tj}F^{\tk}_{\mu\nu} W_{\rho\sigma}
+\frac{4}{3}A^{\ti}A^{\tj}A^{\tk}W_{\mu\nu}W_{\rho\sigma}),
\end{eqnarray}
where $F^{\ti}_{\mu\nu}=2\partial_{[\mu}A^{\ti}_{\nu]}$ are the
Kaluza-Klein-invariant field strengths and $W_{\mu\nu}= 2
\partial_{[\mu}W_{\nu]} $. The kinetic energy term of the scalar
fields  can be written  as \cite{GST1}
$$e^{-1}\mathcal{L}_{S}=-\frac{1}{2}(\tilde{a}_{\ti\tj}\partial_{\mu}\tilde{h}^{\ti}\partial^{\mu}\tilde{h}^{\tj}
+\frac{2}{3}\tilde{a}_{\ti\tj}\partial_{\mu}A^{\ti}\partial^{\mu}A^{\tj}),$$
where we have defined $\tilde{a}_{\ti\tj}=\frac{3}{2}
e^{-2\sigma}\stackrel{\circ}{a}_{\ti\tj},$ and $\tilde{h}^{\ti}=
e^{\sigma} h^{\ti}.$ We note that \[ C_{\ti\tj\tk} \tilde{h}^{\ti} \tilde{h}^{\tj}
\tilde{h}^{\tk} = e^{3\sigma} > 0 . \]

\subsection{Very special K\"{a}hler geometry  and global symmetries}

The above Lagrangian can    be recast   in the language of
 (very) special K\"{a}hler geometry  \cite{4dkahler}
 by  defining the  complex coordinates
        \cite{GST1},
\begin{equation}
z^{\ti}:=\frac{1}{\sqrt{3}}A^{\tI} + \frac{i}{\sqrt{2}}\tilde{h}^{\tI}
\end{equation}
where
\begin{equation}
\tilde{h}^{\ti}:=e^\sigma h^{\ti}. \label{htildeI}
\end{equation}
These $(\tn+1)$ complex coordinates $z^{\ti}$ can be interpreted as
the inhomogeneous coordinates corresponding to  the
$(\tn+2)$-dimensional complex vector \footnote{ We label the top
component of $X^A$ as $\check{0}$ so as to distinguish it from the
zeroth component $X^0$  of $X^{\ti}$ .}
\begin{equation}
X^A=  \left( \begin{array}{c}
X^{\check{0}}\\
X^{\ti}\end{array}  \right) =    \left( \begin{array}{c}
1 \\
z^{\ti}
\end{array} \right)  .
\end{equation}
Introducing the ``prepotential''
\begin{equation}
F(X^A)=-\frac{\sqrt{2}}{3}
C_{\ti\tj\tk}\frac{X^{\tI}X^{\tJ}X^{\tK}}{X^{\check{0}}}
\label{prepot}
\end{equation}
and the symplectic section
\begin{equation}
\left( \begin{array}{c}
X^A\\
F_{A}   \end{array}
\right)  \equiv  \left( \begin{array}{c}
X^A\\
 \partial_{A}F \end{array}
\right)
\end{equation}
one can define a K\"{a}hler potential
\begin{eqnarray}
K(X,\bar{X})&:=&-\ln [i\bar{X}^{A}F_{A}-iX^{A}\bar{F}_{A}] \label{symK}\\
&=&-\ln  \left[  i\frac{\sqrt{2}}{3}C_{\tI\tJ\tK}(z^{\tI}-\bar{z}^{\tI})(z^{\tJ}-\bar{z}^{\tJ})(z^{\tK}-\bar{z}^{\tK})  \right]
\end{eqnarray}
and a ``period matrix''
\begin{equation}
\mathcal{N}_{AB}:=\bar{F}_{AB}+2i\frac{\textrm{Im}(F_{AC})\textrm{Im}(F_{BD})X^{C}X^{D}}{\textrm{Im}(F_{CD})X^{C}X^{D}},
\end{equation}
where $F_{AB}\equiv \partial_{A}\partial_{B}F$ etc.
The particular (``very special'') form (\ref{prepot}) of the prepotential
 leads to
 \begin{equation}\label{metric}
g_{\tI\bar{\tJ}}\equiv \partial_{\tI}\partial_{\bar{\tJ}}K=\frac{3}{2} e^{-2\sigma} {\stackrel{\circ}{a}}_{\tI\tJ}
 \end{equation}
for the K\"{a}hler metric, $g_{\tI\bar{\tJ}}$, on the scalar manifold $\M^{(4)}$  of the four-dimensional theory, and
\begin{eqnarray}
\mathcal{N}_{\check{0}\check{0}}&=&-\frac{2\sqrt{2}}{9\sqrt{3}}C_{\tI\tJ\tK}A^{\tI}A^{\tJ}A^{\tK}
-\frac{i}{3} \left( e^{\sigma}{\stackrel{\circ}{a}}_{\tI\tJ} A^{\tI}A^{\tJ}+\frac{1}{2} e^{3\sigma} \right)\\
\mathcal{N}_{\check{0}\tI}&=&\frac{\sqrt{2}}{3}C_{\tI\tJ\tK}A^{\tJ}A^{\tK}+\frac{i}{\sqrt{3}} e^{\sigma} {\stackrel{\circ}{a}}_{\tI\tJ} A^{\tJ}\\
\mathcal{N}_{\tI\tJ}&=&  -\frac{2\sqrt{2}}{\sqrt{3}}C_{\tI\tJ\tK}A^{\tK}-ie^{\sigma}
{\stackrel{\circ}{a}}_{\tI\tJ} \label{NIJ}
\end{eqnarray}
for the period matrix $\mathcal{N}_{AB}$.
Defining
\begin{equation}
F_{\mu\nu}^{\check{0}}:=-2\sqrt{3}W_{\mu\nu}, \label{WF0}
\end{equation}
  the dimensionally reduced Lagrangian (\ref{redlag}) simplifies to
  \begin{eqnarray}
   e^{-1}\mathcal{L}^{(4)} &=&-\frac{1}{2}R  -
   g_{\ti\bar{\tJ}}    (\partial_{\mu}z^{\ti})(\partial^{\mu} \bar{z}^{\tJ})
   \nonumber \\
   &&+\frac{1}{4}\textrm{Im}(\mathcal{N}_{AB})F_{\mu\nu}^{A}F^{\mu\nu B}-\frac{1}{8}
   \textrm{Re} (\mathcal{N}_{AB})\epsilon^{\mu\nu\rho\sigma}
   F_{\mu\nu}^{A}F_{\rho\sigma}^{B}.\label{redlag1b}
   \end{eqnarray}

The scalar fields $z^{\tI}$ are restricted to the domain  $
\mathcal{V}(Im(z)) >0$. As was shown in \cite{GST1}, the scalar
manifolds of    the  4D, $\mathcal{N}=2$   MESGTs obtained by
dimensional reduction from five dimensions  are simply the
generalized ``upper half planes" with respect to the cubic form for
all the MESGTs. For those theories defined by Euclidean Jordan
algebras of degree 3 they are the K\"ocher upper half planes of the
corresponding Jordan algebras. These are sometimes known as Siegel
domains of the first kind. Cecotti showed that  the more general
class  of homogeneous (not necessarily symmetric) scalar manifolds
of   4D MESGTs are also Siegel domains of the first kind
\cite{cecotti}.   The upper half planes of the Jordan algebras can
be mapped into bounded symmetric domain, which can be realized as
hermitian symmetric spaces. For the Jordan algebras of degree 3, we
list the scalar manifolds of the corresponding   4D  MESGTs
describing the coupling of $(\tn +1)$ vector multiplets to
supergravity in the table below \cite{GST1}:

\begin{eqnarray}\label{4dmagicals}
J_{3}^{\mathbb{R}}:\quad \mathcal{M}&=& Sp(6,\mathbb{R})/ U(3)\qquad
\qquad \qquad \quad
(\tn +1=6)\nonumber\\
J_{3}^{\mathbb{C}}:\quad \mathcal{M}&=& SU(3,3)/ S(U(3)\times U(3))
\qquad \,
(\tn +1=9)\nonumber\\
J_{3}^{\mathbb{H}}:\quad \mathcal{M}&=& SO^{*}(12)/ U(6)\qquad
\qquad \qquad \quad
(\tn +1=15) \nonumber \\
J_{3}^{\mathbb{O}}:\quad \mathcal{M}&=& E_{7(-25)}/ E_6 \times U(1)
\qquad \qquad  \quad (\tn +1=27) \nonumber \\
J= (\mathbb{R} \oplus \Sigma_{\tn}): \quad \mathcal{M}& =&
\frac{SO(\tn,2)\times SO(2,1)}{SO(\tn) \times SO(2) \times SO(2)}
 \quad \quad \, (\tn +1)  \label{symlist}
\end{eqnarray}

In \cite{dwvp92}, isometries of  general homogeneous manifolds of
very special real,  K\"{a}hler and quaternionic type were studied.
More specifically , the authors of \cite{dwvp92} first classified
very special real scalar manifolds of 5D, $\mathcal{N}=2$ MESGTs
that are homogeneous , but not necessarily symmetric. They then
studied the isometries of very  special K\"{a}hler and quaternionic
spaces that follow from the dimensional reduction of those $5D$
theories to four and three dimensions, respectively. They confirmed
and generalized the earlier results showing that there is a map,
\textbf{r}, from very special real manifolds of dimension $\tn$ to very
special K\"{a}hler manifolds of complex dimension $\tn+1$, and a map,
\textbf{c}, from special K\"{a}hler manifolds of complex dimension
$\tn+1$ to special quaternionic manifolds of quaternionic dimension
$\tn+2$. This phenomenon was first observed for theories originating
from   5D   MESGTs defined by Jordan algebras of degree 3
with symmetric target spaces \cite{GST1}.

Upon dimensional reduction from five dimensions, the resulting Lagrangian is still
 invariant under the action
of the symmetry group, $G$, of $C_{\ti\tj\tk}$. Its action on the
complex scalars $z^{\ti}$  parameterizing the K\"{a}hler manifold is
simply
$$\delta z^{\ti} = B^{\ti}{}_{\tj}z^{\tj},$$ which leaves the K\"{a}hler potential
$K(z,\bar{z})=-\ln \mathcal{V}(z-\bar{z}) + \textrm{const.} ,$ invariant.
The vector fields  $A^{A}_{\mu}$ transform in  a reducible
representation under $G$, with the components transforming as
$$\delta A^{\ti}_{\mu}=B^{\ti}{}_{\tj} A^{\tj}_{\mu}$$
$$\delta W_{\mu}=0.$$

Under shifts of the scalars $A^{\ti} \rightarrow A^{\ti} + b^{\ti}$,
which come from the vector fields in five dimensions, the K\"ahler
potential is manifestly invariant. Furthermore, one finds that under
\emph{real constant} shifts of the scalar fields
\[
z^{\ti}\rightarrow z^{\ti}+b^{\ti}, \] together with the shifts
\[ F^{\ti}_{\mu\nu}\rightarrow F^{\ti}_{\mu\nu}
- 2W_{\mu\nu}b^{\ti}
\]
 the reduced Lagrangian is also invariant up to surface terms which
are interpreted as shifts of the theta angle of the $4D$
supersymmetric theory.  Furthermore, for any dimensionally reduced
 supergravity theory,  there is
 also a scaling symmetry $SO(1,1)$, whose  generator is denoted as $\beta$.
It acts as $X^{\check{0}}  \rightarrow             e^{\beta}
X^{\check{0}} , \quad
 X^{\ti}   \rightarrow   e^{-\frac{\beta}{3}}  X^{\ti}  $.

The authors of \cite{dwvp92} also elucidated the conditions under
which the additional, ``hidden", isometries, with infinitesimal
parameters $a^{\ti}$, exist which do not have a clear origin from
the dimensional reduction of the $5D$ theories.  The number of
hidden isometries is determined by the maximal number of non-zero
parameters in ${a_0,\ldots ,a_n}$ that satisfy \eq
a_{\tM}E^{\tM}_{\ti\tj\tk\tl}=0, \en where
\[ E^{\tM}_{\ti\tj\tk\tl}:=\frac{-27}{(\mathcal{V}(\tilde{h}))^2}C^{\tE\tF\tM}
C_{\tE(\ti     \tj}C_{\tk\tl)\tF}
 - \delta^{\tM}_{(\ti} C_{\tj\tk\tl)} \]
 and
\begin{eqnarray}
C^{\tilde{E}\tilde{F}\tilde{M}}&\equiv&
{\stackrel{\circ}{a}}^{\tilde{E}\ti}
 {\stackrel{\circ}{a}}^{\tilde{F}\tj} {\stackrel{\circ}{a}}^{\tilde{M}\tk}
C_{\ti\tj\tk}\\
  \mathcal{V}(\tilde{h}) &\equiv& C_{\ti\tj\tk} \tilde{h}^{\ti} \tilde{h}^{\tj} \tilde{h}^{\tk} =
  e^{3\sigma}.
\end{eqnarray}

 We should note that whereas $C_{\ti\tj\tk}$ are constant, $C^{\ti\tj\tk}$ are scalar field dependent in general.
  As was shown in \cite{dwvp92}
 all the isometries of the
 scalar manifold extend to the full generalized duality group of the MESGT in
  four dimensions.

The Lie algebra of the full set of  isometries has a  three-graded
decomposition with respect to the eigenvalues of the scale symmetry
generator $\beta$ :
\begin{equation}
\mathcal{W}=\mathcal{W}^{-1}\oplus\mathcal{W}^{0}\oplus\mathcal{W}^{+1}, \label{Walgebra}
\end{equation}
where $\mathcal{W}^{0}$ consists of the generators of
$B^{\ti}{}_{\tj}$ and $\beta$, $\mathcal{W}^{-1}$ is associated with
the real parameters $a^{\ti}$, and $\mathcal{W}^{+1}$ is associated
with the real parameters $b^{\ti}$.

It was originally  stated  in \cite{cremmeretal} that the subalgebra
$$\mathcal{W}^{0}\oplus\mathcal{W}^{+1}$$ is the maximal symmetry subalgebra of the full duality group
under which the  Lagrangian is invariant and, hence, to obtain
$\mathcal{N}=2$ YMESGTs one may then  try   to gauge subgroups of
this parabolic subgroup. However this statement is  true only for
the symplectic section of the theory that comes from direct
dimensional reduction      \cite{dwvp92}.
 For
example, for symmetric scalar manifolds the full duality group has
negative grade generators and, via dualizations of the vector
fields,  one can go to different  symplectic sections of the
respective $\mathcal{N}=2$ MESGT such that the Lagrangian becomes
invariant under other subgroups of the full duality group generated
by $\mathcal{W}$ that are not subgroups of the parabolic group
generated by $\mathcal{W}^{0}\oplus\mathcal{W}^{+1}$ as will be
discussed in the next section.

The Lie algebras $\mathcal{W}$ of the isometries of the symmetric
scalar manifolds of four-dimensional theories obtained from those
   5D  theories defined by Euclidean Jordan algebras of degree
three can be identified with the Lie algebras of the conformal
(or ``linear fractional'') groups of the corresponding Jordan algebras. As
explained in appendix A, the conformal groups of Jordan algebras are
generated by translations, $\mathfrak{T}_J$, special conformal
generators, $\mathfrak{K}_J$, and the generators of the structure
group (which is simply the direct product of the generalized
     ``Lorentz group''   of the Jordan algebra and dilatations, as explained in Appendix A). The
conformal Lie algebra of $J$ has a 3-graded decomposition with
respect its structure algebra

\[ \mathfrak{\textrm{Conf}}(J) = \mathfrak{T}_J \oplus \mathfrak{str}(J) \oplus  \mathfrak{K}_J ,\]
which coincides with the three grading of $\mathcal{W}$ (eq.
(\ref{Walgebra})) by identification of the dilatation generator with
$\beta$. The K\"ocher upper half-space $(J+iJ_+)$ of a Jordan
algebra is spanned by complexified elements of the form  $ X+iY$
where $X,Y\in J$ such that  $Y$ has positive norm. The upper
half-space of a Euclidean Jordan algebra $J$ can be mapped into the
Hermitian symmetric space \cite{koecher2}
\[ (J+iJ_+)\;\;\Longleftrightarrow\;\;\frac{\textrm{Conf}(J)}{K(J)}, \]
where $K(J)$ is the subgroup generated by the derivations
(``rotations'') and generators of the form $(
\mathfrak{T}_J-\mathfrak{K}_J )$. For Euclidean Jordan algebras
$K(J)$ turns out to be the maximal compact subgroup of $\textrm{Conf}(J)$.
For the MESGTs defined by       Euclidean Jordan algebras of degree three,
  the corresponding symmetric      spaces  were listed in (\ref{symlist}).

 The scalar manifolds, $\mathcal{M}_4$, of the four-dimensional
 theories obtained by dimensionally reducing a    5D   unified MESGT
defined by Lorentzian Jordan algebras are, in general,  the  ``upper
half-spaces of  cubic hypersurfaces" defined by $d_{\ti\tj\tk}$. The
complex coordinates of $\mathcal{M}_4$ can be written as
\[ z^{\ti}= \sqrt{\frac{1}{2}}(\sqrt{\frac{2}{3}}A^{\ti} + i \tilde{h}^{\ti} ) \]
such that
\[ d_{\ti\tj\tk} \tilde{h}^{\ti} \tilde{h}^{\tj} \tilde{h}^{\tk} > 0 .\]
with its metric given by equation (\ref{metric}). Formally we can
write
\[
\mathcal{M}_4\;\;\Longleftrightarrow\;\; A + i \tilde{h} ,
\]
where $A=e_{\ti} A^{\ti}$ and $\tilde{h} =e_{\ti} \tilde{h}^{\ti}$
are traceless elements of the Lorentzian Jordan algebra such that
$A$ is unconstrained otherwise  and $\tilde{h}$ has positive cubic
norm. From this it is evident that the scalar manifold
$\mathcal{M}_4$ can be mapped into a subspace of the Koecher upper
half space of the full Jordan algebra $J_{(1,N)}^{\mathbb{A}}$
spanned by elements of the form $X+iY$ where $ X$ and $Y$ are
general elements (not necessarily traceless) of the Jordan algebra
$J_{(1,N)}^{\mathbb{A}}$ such that the standard norm
$\mathcal{P}(Y)$ of degree $N+1$ of the element $Y$ is positive. The
upper half-planes of the Jordan algebras $J_{(1,N)}^{\mathbb{A}}$,
with an appropriately defined metric, can be identified with the
coset spaces of the form
\[
\frac{\textrm{Conf}(J_{(1,N)}^{\mathbb{A}})}{\textrm{K}(J_{(1,N)}^{\mathbb{A}})}
.
\]
 In Table 2 we give the list of the groups that underlie these spaces for all Lorentzian Jordan algebras
 $J_{(1,N)}^{\mathbb{A}}$.

 \begin{table}[ht]
\begin{center}
\begin{displaymath}
\begin{array}{|c|c|c|}
\hline
~&~&~\\
J & \textrm{Conf}(J) & \textrm{K}(J)  \\
\hline
~&~&~\\
J_{(1,N)}^{\mathbb R}& Sp(2N+2,\mathbb{R}) & SU(N,1) \times U(1) \\
~&~&~\\
J_{(1,N)}^{\mathbb C} & SU(N+1,N+1) & SU(N,1)\times SU(N,1) \times U(1) \\
~&~&~\\
J_{(1,N)}^{\mathbb H} &SO^*(4N+4) & SU(2N,2) \times U(1)   \\
~&~&~\\
J_{(1,2)}^{\mathbb O} & E_{7(-25)} & E_{6(-14)} \times U(1)  \\
 \hline
\end{array}
\end{displaymath}
\end{center}

\caption{List of the simple Lorentzian Jordan algebras of type
$J_{(1,N)}^{\mathbb{A}}$ and their conformal (M\"{o}bius) groups and
their subgroups $K(J)$.}
\end{table}

 As was discussed in section 2, the
cubic form defined over the traceless elements of the Lorentzian
Jordan algebras $J_{(1,3)}^{\mathbb{R}}, J_{(1,3)}^{\mathbb{C}} $
and $J_{(1,3)}^{\mathbb{H}}$ coincide with the generic cubic norms
of degree three Euclidean Jordan algebras $J_3^{\mathbb{C}},
J_3^{\mathbb{H}}$ and $J_3^{\mathbb{O}}$, respectively. Hence for
the $\mathcal{N}=2$ MESGTs defined by $J_{(1,3)}^{\mathbb{R}},
J_{(1,3)}^{\mathbb{C}} $ and $J_{(1,3)}^{\mathbb{H}}$ the scalar
manifolds are the symmetric spaces listed in Table 3.

\begin{table}[ht]
\begin{center}
\begin{displaymath}
\begin{array}{|c|c|c|}
\hline
~&~&~\\
J & Aut(J)\circledS T_J & \mathcal{M}_4  \\
\hline
~&~&~\\
J_{(1,3)}^{\mathbb R}& SO(3,1)\circledS T_9 & Sp(6,\mathbb{R})/U(3) \\
~&~&~\\
J_{(1,3)}^{\mathbb C} & SU(3,1)\circledS T_{15} & SO^*(12)/U(6) \\
~&~&~\\
J_{(1,3)}^{\mathbb H} &USp(6,2) \circledS T_{27} & E_{7(-25)}/E_6 \times U(1) \\
~&~&~ \\

\hline
\end{array}
\end{displaymath}
\end{center}

\caption{ Simple Lorentzian Jordan algebras of type
$J_{(1,3)}^{\mathbb{A}}$          for $\mathbb{A}=\mathbb{R}, \mathbb{C},\mathbb{H}$. The second  column lists the manifest
isometries and third column lists the scalar manifolds of
$\mathcal{N}=2$ MESGTs in 4D defined by them.}
\end{table}

\section{Unified    MESGTs   and YMESGTs in Four Dimensions}

In this section, we will give a classification of   four-dimensional
unified $\mathcal{N}=2$ MESGTs and YMESGTs. Since a full
classification of four-dimensional MESGTs with non-homogeneous
scalar manifolds and non-trivial isometry groups does not yet exist
we will restrict our classification to theories with homogeneous
scalar manifolds and those theories with non-homogeneous manifolds
defined by Lorentzian Jordan algebras.   As mentioned earlier,   the
  analysis in four
dimensions is more complicated than in five due to the fact that the
full U-duality group, $\mathcal{U}$, is an on-shell symmetry and not
the symmetry of the Lagrangian in four dimensions. The electric
field strengths $\mathcal{F}^{A\;\mu\nu}$ together with the magnetic
(dual)
field strengths $\mathcal{G}_{A\;\mu\nu}$ form a representation of
the U-duality group $\mathcal{U}$, which may be reducible. We shall
call a MESGT in 4D unified if the electric field strengths
$\mathcal{F}^{A\;\mu\nu}$, including that of the graviphoton, form
an irreducible representation of a simple symmetry group $G$ of the
Lagrangian.  Since, in 4D, massless vector fields are dual to
other vector fields, we have some freedom in choosing electric field
strengths versus the magnetic field strengths. By dualizing some of
the vector fields  one can obtain some {\it real} linear
combinations of the $ ( \mathcal{F}^{A\;\mu\nu}$ and $
*(\mathcal{G})_{A\;\mu\nu} )$ to be the new electric field
strengths. In so doing one also changes the subgroup $G$ of the full
U-duality group that is a symmetry of the Lagrangian. This
phenomenon can best be explained by the example of maximal
supergravity. The five-dimensional  $\mathcal{N}=8$ supergravity
with   U-duality group $E_{6(6)}$ yields, upon dimensional reduction
to   four dimensions, the parabolic group
  \[ [E_{6(6)}\times
SO(1,1)] \circledS T_{27} \] as the symmetry of the Lagrangian
\footnote{ The symbol $\circledS$  denotes semidirect product and
$T_{27}$ denotes the translation group in 27 dimensions.}. However
we know that by dualization one can obtain $SL(8,\mathbb{R})$
\cite{cjs} or $SU^*(8)$ \cite{grw,hull} as the symmetry of the
Lagrangian. We should stress that both $SL(8,\mathbb{R})$ and
$SU^*(8)$ are subgroups of the full U-duality group $E_{7(7)}$ under
which electric and magnetic fields transform in $\mathbf{28}$ and
$\mathbf{\tilde{28}}$, which are real representations of
$SL(8,\mathbb{R})$ and $SU^*(8)$, respectively. Even though $SU(8)$
is also a subgroup of $E_{7(7)}$ under which
$\mathbf{56}=\mathbf{28} \oplus \mathbf{\overline{28}}$, it can not
be made a symmetry of the Lagrangian by dualization, since 28 is a
complex representation of $SU(8)$ and would thus  require complex
vector potentials.

One should  also note that,  in contrast to the
situation in five dimensions, neither the irreducibility of the
representation $ ( \mathcal{F}^{A\;\mu\nu} \oplus
\mathcal{G}_{A\;\mu\nu} )$ under $\mathcal{U}$ nor the simplicity of
the group $\mathcal{U}$ is necessary or sufficient for having a
unified MESGT in four dimensions.
 This will also be illustrated in some of our examples later in the text.

We will begin our analysis by studying the symmetries of \emph{pure}
5D, $\mathcal{N}=2$ supergravity theories dimensionally reduced
to four dimensions and its gaugings.

\subsection{ Pure   5D, $\mathcal{N}=2$
 supergravity dimensionally
reduced to 4D.} Since the symmetry group structure of dimensionally
reduced pure supergravity will be helpful in understanding the
action of U-duality group in the general case we shall review it
first.  The dimensional reduction of pure 5D, $\mathcal{N}=2$
supergravity to 4D was first studied long time ago in
\cite{shamnico} and more recently in \cite{mizo}. The resulting
theory is $\mathcal{N}=2$, 4D supergravity coupled to one vector
multiplet and the scalar manifold is $SU(1,1)/U(1)$, where the two
scalars come from the graviton and the graviphoton in five
dimensions. The manifest symmetry of the dimensionally reduced
Lagrangian is $SO(1,1)_{\beta}\circledS T$ where $T$ represents  the
translations of the scalar coming from the graviphoton in five
dimensions and the generator of scale transformations
$SO(1,1)_{\beta}$ is $\beta$. The ``hidden''  isometry of the
reduced theory corresponds to special conformal transformations $K$
that extends $SO(1,1)_{\beta}\circledS T$ to $SU(1,1)$. The compact
$U(1)$ generator of $SU(1,1)$ is $\mathfrak{t}-\mathfrak{k}$, where
$\mathfrak{t}$ and $\mathfrak{k}$ are generators of translations and
special conformal transformations, respectively. The $SU(1,1)$ acts
as the full U-duality group on-shell. Remarkably, the two vector
field strengths and their duals transform in the irreducible
four-dimensional representation of $SU(1,1)$ \cite{mizo}  (In the
language of compact $SU(2)$ it is the spin $3/2$ representation of
$SU(1,1)$). What this means is that all the field strengths and
their duals carry a charge under any Abelian subgroup of $SU(1,1)$,
be it $U(1)$ or $SO(1,1)$. Since the gauge field of an Abelian gauge
theory must be neutral under the Abelian group, one can not gauge
any Abelian subgroup of $SU(1,1)$.  We shall denote the full
U-duality group of dimensionally reduced pure $\mathcal{N}=2$
supergravity  as $SU(1,1)_G$. All other $\mathcal{N}=2$ MESGTs
dimensionally reduced to 4D can be truncated to this theory.
$SU(1,1)_G$ will be a subgroup of the full U-duality groups of
dimensionally reduced MESGTs  defined by Euclidean Jordan algebras
of degree three , whose scalar manifolds are symmetric spaces.
However, in general, only the $SO(1,1) \circledS T$ subgroup of
$SU(1,1)_G$ will be a symmetry of dimensionally reduced MESGTs as
will become evident when we discuss theories whose scalar manifolds
are homogeneous but not symmetric.

\subsection{Theories whose scalar manifolds are symmetric
spaces}
\begin{itemize}
\item
The generic Jordan family with target manifolds
\[ \frac{SO(\tn-1,1)}{SO(\tn-1)}\times SO(1,1) \] in 5D lead to four
dimensional MESGTs with target manifolds \cite{GST1}

\eq \mathcal{M}= \frac{SO(\tn,2)}{SO(\tn)\times SO(2)} \times \frac{SU(1,1)}{U(1)} \en

Under the U-duality group, $\mathcal{U}= SO(\tn,2)\times SU(1,1)$,
the   ($\tn+2$) field strengths $\mathcal{F}^{A\;\mu\nu}$ and their
magnetic duals $\mathcal{G}_{A\;\mu\nu}$ transform in the $(\tn+2,
2)$ representation. One can take a symplectic section of these
theories such that all the electric field strengths
$\mathcal{F}^{A\;\mu\nu}$ transform in the real irreducible ($\tn
+2$) representation of the simple subgroup $SO(\tn,2)$ of
$\mathcal{U}$. Thus the generic Jordan family of $\mathcal{N}=2$
MESGTs in 4D are unified theories, in contrast to their
five-dimensional counterparts. To obtain a unified YMESGT by gauging
a simple subgroup $H$ of $SO(\tn,2)$, all the $(\tn+2)$ electric
field strengths must transform in the adjoint representation of $H$.
This is only possible for $\tn=1$. Thus the unique unified YMESGT in
the generic Jordan family has $SO(2,1)$ as its simple gauge group
\footnote{Interestingly, this unified YMESGT forms a common
subsector of several  of the models with stable de Sitter vacua
considered in \cite{FTVP}. However, it is really  only a subsector,
and additional vector multiplets and gauge group factors are
necessary in order to obtain
 stable de Sitter vacua \cite{FTVP}. Clearly, with these additional gauge group factors, these theories as a whole  are then no longer unified YMESGTs.}.

We should stress that the $SU(1,1)=SO(2,1)$ factor in the four
dimensional U-duality group is not the $SU(1,1)_G$ symmetry of the
pure supergravity dimensionally reduced to four dimensions. This is
evident from the fact that the field strengths and their duals form
2 dimensional (spin $1/2$) representations of the former. The
$SU(1,1)_G$ is a diagonal subgroup of $SU(1,1)$ factor and an
$SO(2,1)$ subgroup of $SO(\tn,2)$ under which we have the following
decompositions:

\[ SO(2,1) \times SO(\tn,2) \supset SO(2,1) \times SO(2,1) \times
SO(\tn -1) \supset SU(1,1)_G \times SO(\tn -1) \]

\[ (2,\tn+2) = (2,3,1) \oplus (2,1,\tn -1) = (4,1) \oplus (2,1)
\oplus (2,\tn -1) \]

Note that the centralizer of $SU(1,1)_G$ inside the four-dimensional
U-duality group is the automorphism group of the corresponding
Jordan algebra, which is $SO(\tn-1)$. This is a general feature of
all theories defined by Jordan algebras of degree three as will be
made evident below.

\end{itemize}
Let us next analyze the magical supergravity theories defined by
\emph{simple} Jordan algebras of degree 3, whose scalar manifolds are all
symmetric.
\begin{itemize}
\item
 For the $\mathcal{N}=2$ MESGT defined by
$J_3^{\mathbb{R}}$, the four-dimensional U-duality group
  is $Sp(6,\mathbb{R})$,
which is also the isometry group of its scalar manifold,

\eq \mathcal{M} = \frac{Sp(6,\mathbb{R})}{U(3)}. \en

Under the U-duality group $Sp(6,\mathbb{R})$,   the $(\tn+2)=7$
electric field strengths of this theory together with their magnetic
duals transform in the irreducible   14-dimensional  representation.
However, $Sp(6,\mathbb{R})$ does not have a simple subgroup $G$
under which all the seven electric field strengths transform
irreducibly. Thus the four-dimensional $\mathcal{N}=2$ theory
defined by $J_3^{\mathbb{R}}$ is \emph{not }a unified MESGT. This
provides an example of a theory with a simple U-duality group,
$\mathcal{U}$, under which electric and magnetic field strengths
transform in an irrep of $\mathcal{U}$, but which is nevertheless
 not unified by our definition.  Consequently,  it
is also  \emph{not } possible to gauge this theory so as to obtain a
unified YMESGT.

The centralizer of $SU(1,1)_G$ inside the four-dimensional U-duality
group $Sp(6,\mathbb{R})$ is $SO(3)$ which is the automorphism group
of $J_3^{\mathbb{R}}$ :
\[ Sp(6,\mathbb{R}) \supset SU(1,1)_G \times SO(3) \]
\[ 14 = (4,1) + (2,5) \]
\item
The MESGT defined by the Jordan algebra $J_3^{\mathbb{C}}$ has the
U-duality group $SU(3,3)$ in 4D, under which the field strengths
and their magnetic duals, $\mathcal{F}^{A\;\mu\nu} \oplus
\mathcal{G}_{A\;\mu\nu}$, transform in the irreducible symplectic
representation $\mathbf{20}$.  The group $SU(3,3)$ has a subgroup
under which the $20$ decomposes into two real ten-dimensional irreps, namely,  $SO(3,3) \cong SL(4,\mathbb{R})$ \footnote{ Under the
$SU(3,2)$ subgroup 20 also decomposes as $10 \oplus \bar{10}$ .
However they are complex. }
\[ SU(3,3) \supset SL(4,\mathbb{R}) \]
\[ 20 = 10 + \tilde{10}. \]
Therefore, one can take a symplectic section of this theory such that
$SL(4,\mathbb{R})$ becomes the electric subgroup of the full
U-duality group. Hence this theory defined by $J_3^{\mathbb{C}}$ is
a unified MESGT. Furthermore, one can gauge this theory to obtain a
unified YMESGT with the gauge group $SO(3,2) \cong
Sp(4,\mathbb{R})$, since under the $SO(3,2)$ subgroup all the
electric field strengths transform in the adjoint representation.

The centralizer of $SU(1,1)_G$ inside $SU(3,3)$ is $SU(3)$
\[ SU(3,3) \supset SU(1,1)_G \times SU(3) \]
\[ 20= (4,1) +(2,8) \]
\item
The theory defined by the quaternionic Jordan algebra
$J_3^{\mathbb{H}}$ has the U-duality group $SO^*(12)$ in four
dimensions under which the field strengths of the vector fields
together with their duals transform in the spinor representation
$\mathbf{32}$. The maximal rank subgroups of  $SO^*(12)$ are

\begin{eqnarray}
SO^*(12) & \supset &  SO(6,\mathbb{C}) \cong SL(4,\mathbb{C}) \nonumber \\
SO^*(12) & \supset &  SO^*(2p) \times SO^*(12-2p) \quad , p=1,2,3  \nonumber \\
SO^*(12) & \supset & U(6) \\
SO^*(12) & \supset &  SU(p,6-p) \times U(1) \nonumber \\
SO^*(12) & \supset & SU^*(6) \times SO(1,1) \nonumber
\end{eqnarray}

Under the $SO^*(10) \times SO(2)$ subgroup the spinor $\mathbf{32}$
of $SO^*(12)$ decomposes as $\mathbf{16}^{+1} \oplus
\mathbf{\overline{16}}^{-1}$. The spinor representation $16$ of
$SO^*(10)$ is complex \footnote{$SO^*(10)$ does not admit any
Majorana-Weyl spinors.}. However, under the subgroup
$SO(6,\mathbb{C}) \cong SL(4,\mathbb{C})$, the spinor
representation $32$ of $SO^*(12)$ decomposes as
\[ 32 = (4,\bar{4}) \oplus (\bar{4},4) \]
Since the $(4,\bar{4})$ representation of $SL(4,\mathbb{C})$ is
real, the corresponding MESGT is a unified theory. The only simple
noncompact group of dimension 16 is $SL(3,\mathbb{C})$, which is a
subgroup of $SL(4,\mathbb{C})$. However, $(4,\bar{4})$ of
$SL(4,\mathbb{C})$ does not transform irreducibly under the
$SL(3,\mathbb{C})$ subgroup and hence this theory can not be gauged
so as to obtain a unified YMESGT in 4D.

We should recall that the corresponding theory in five dimensions is
a unified MESGT and can be gauged \cite{GST2} so as to obtain the
unique unified YMESGT with gauge group $SO^*(6)=SU(3,1)$ in 5D
whose scalar manifold is a symmetric space.

For this four-dimensional theory the centralizer of $SU(1,1)_G$ is
$USp(6)$
\[ SO^*(12) \supset SU(1,1) \times Usp(6) \]
\[ 32 = (4,1) + (2,14) \]
\item
The exceptional supergravity theory defined by the Jordan algebra
$J_3^{\mathbb{O}}$ in four dimensions has the U-duality group
$E_{7(-25)}$ with the maximal compact subgroup $E_6\times U(1)$. The
28 electric field strengths together with their magnetic duals
transform in the 56 dimensional irrep of $E_{7(-25)}$. The
interesting subgroups of $E_{7(-25)}$ are
\begin{eqnarray}
E_{7(-25)} &\supset& SU^*(8) \nonumber \\
E_{7(-25)} &\supset& SU(6,2) \nonumber \\
E_{7(-25)} &\supset& SO(10,2) \times SO(2,1) \\
E_{7(-25)} &\supset& SO^*(12) \times SU(2)  \nonumber \\
 E_{7(-25)} &\supset& E_{6(-26)} \times SO(1,1)   .     \nonumber
\end{eqnarray}
Under $SO(10,2)\times SO(2,1)$ and $SO^*(12) \times SU(2)$ the 56 of
$ E_{7(-25)}$ decomposes as
\[ 56=  (12,2) + (32,1), \]
and hence $SO(10,2)$ or $SO^*(12)$ cannot act irreducibly on all
the electric field strengths. The same holds  true for $E_{6(-26)}$
under which the 56 of $E_{7(-25)}$ decomposes as \[ 56 =1+27 +
\overline{27} + 1 \]

On the other hand, under the $SU^*(8)$ or $SU(6,2)$ subgroup the 56
of $E_{7(-25)}$ decomposes as \[ 56=28 + \tilde{28}. \] Now, the
  28-dimensional  representation of $SU(6,2)$ is complex and hence
$SU(6,2)$ can not be a symmetry of the Lagrangian. On the other hand,
the $28$ is a real representation of $SU^*(8)$ and one can choose a
symplectic section of the theory via dualization such that $SU^*(8)$
becomes  a symmetry of the Lagrangian with
 all the electric field strengths
transforming irreducibly in the 28. Thus, the exceptional MESGT is a
unified   MESGT.  Furthermore, one can gauge the $SO^*(8)=SO(6,2) =
SU^*(8) $ subgroup so as to obtain a unified YMESGT.
$SO^*(8)=SO(6,2) $  is the common maximal subgroup of $SU(6,2)$ and
$SU^*(8)$.

 We should  note the important fact that the $SO(6,2)$ gauged exceptional
supergravity is the $ \mathcal{N}=2$ supersymmetric analog of the
$SO(6,2)$ gauged maximal $\mathcal{N}=8$ supergravity
\cite{hullso62},  where $SO(6,2)$ is the common subgroup of the
$SU(6,2)$ and $SU^*(8)$ subgroups of $E_{7(7)}$. This extends
further the parallels between the exceptional $ \mathcal{N}=2$
supergravity and the maximal supergravity. They both have different
real forms of $E_6,E_7$ and $E_8$  as their U-duality groups in 5, 4
and 3 dimensions, respectively. Both theories can be gauged in five
dimensions with the noncompact gauge group $SU(3,1)$ and twelve
tensor fields \cite{grw2,GZ1}.  They both can be gauged in four
dimensions with the gauge group $SO(6,2)$.

For the exceptional theory the centralizer of $SU(1,1)_G$ is $F_4$
\[ E_{7(-25)} \supset SU(1,1)_G \times F_4 \]
\[ 56 = (4,1) + (2,26) \]

We summarize our results for the symmetric space theories in Table 4.

 \begin{table}[ht]
\begin{center}
\begin{displaymath}
\begin{array}{|c|c|c|}
\hline
~&~&~\\
\mathcal{M} & \textrm{Unified global symmetry} & \textrm{Unified local symmetry}  \\
\hline
~&~&~\\
\frac{SO(\tn,2)}{SO(\tn)\times SO(2)} \times \frac{SU(1,1)}{U(1)}& SO(\tn,2) &
    SO(1,2) \quad (\tn=1) \\
~&~&~\\
\frac{Sp(6,\mathbb{R})}{U(3)} & -- & -- \\
~&~&~\\
\frac{SU(3,3)}{S(U(3)\times U(3))} &SL(4,\mathbb{R} )& SO(3,2)  \\
~&~&~\\
\frac{SO^{\ast}(12)}{U(6)}    &     SL(4,\mathbb{C}) & --      \\
~&~&~\\
\frac{E_{7(-25)}}{E_{6} \times U(1)} & SU^{\ast}(8) & SO(6,2)  \\
~&~&~\\
 \hline
\end{array}
\end{displaymath}
\end{center}

\caption{List of the  very special   4D MESGTs with symmetric target
spaces. The second column displays the possible unified global
symmetries, and the third column lists the possible unified local
symmetry groups.  }
\end{table}

\subsection{Theories with homogeneous, but     non-symmetric  scalar
manifolds}  The generic non-Jordan family of MESGTs in 5D have
scalar manifolds of the form
\[ \frac{SO(\tn,1)}{SO(\tn)} \]
describing the coupling of $\tn$  vector multiplets to
$\mathcal{N}=2$ supergravity \cite{GST3}. As was shown in
\cite{dWvP2}, the corresponding $C_{\ti\tj\tk}$ tensor, and hence
the Lagrangian is invariant only under the subgroup
$[SO(\tn-1)\times SO(1,1)_{\lambda}] \circledS T_{(\tn-1)}$ of
$SO(\tn,1)$. Since, under the action of the maximal simple subgroup
$SO(\tn-1)$, two of the vectors transform as singlets,  the generic
non-Jordan family of MESGTs are not unified in 5D.

As a consequence of the fact that the full isometry group
$SO(\tn-1)$ is broken to its subgroup $[SO(\tn-1)\times SO(1,1)]
\circledS T_{(\tn-1)}$ by the interaction terms of the 5D
Lagrangian, the scalar manifolds of the corresponding dimensionally
reduced MESGTs in 4D are homogeneous, but not symmetric spaces. The
Lie algebras of their isometry groups are not simple and have the
following non semisimple  Lie algebra graded with respect to the
$SO(1,1)_{\alpha}$ generator   $ \alpha= \frac{2}{3} \lambda + \beta
$ \cite{dwvp92}
\[
\mathfrak{g}= \mathfrak{g}^0 \oplus \mathfrak{g}^1 \oplus
\mathfrak{g}^2
\]

where \[\mathfrak{g}^0= \alpha \oplus \mathfrak{so}(2,1)\oplus
\mathfrak{so}(\tn -1) \]and  \[ \mathfrak{g}^1 = (1,2,\tn-1) \]
\[\mathfrak{g}^2 =(2,0,0) \]

The semisimple subgroup of the isometry group  is $SO(2,1) \times
SO(\tn-1)$ under which the $\tn +2$ electric field strengths
decompose as $(3,1) \oplus (1,\tn-1) $. Thus the corresponding
MESGT's are not unified.

Let us now consider the $4D$ MESGTs that are obtained from $5d$
theories whose scalar manifolds (in five dimensions) are homogeneous, but
not symmetric.
  A complete list of possible scalar manifolds of $5d$ MESGTs that are  homogeneous spaces was given in \cite{dWvP1}.
  To achieve this  classification the authors of \cite{dWvP1}
showed  that the requirement of
 a transitive isometry group allows one to transform  the most general
solution for the symmetric tensor from the ``canonical basis''
 to the basis:
\begin{eqnarray}
C_{011}=1 \\ \nonumber C_{0\bar{\mu}\bar{\nu}}= - \delta_{\bar{\mu}\bar{\nu}} \\ \nonumber C_{1\bar{i}\bar{j}}= -
\delta_{\bar{i}\bar{j}} \\ \nonumber C_{\bar{\mu}\bar{i}\bar{j}}=\gamma_{\bar{\mu}\bar{i}\bar{j}}
\end{eqnarray}
where the indices $\ti$ are now split such that
$\ti=0,1,\bar{\mu},\bar{i}$ with $\bar{\mu}=1,2,..,q+1 $ and
$\bar{i}=1,2,..,r$ ($\tn=r+q+2$) . The coefficients
 $\gamma_{\bar{\mu}\bar{i}\bar{j}}$ are ($q+1$) real $r\times r$ matrices
that generate a real Clifford algebra, ${\cal C}(q+1,0)$, of
positive signature. The allowed homogeneous (but not symmetric)
spaces are, in general, quotients of  ``parabolic groups'' $G$
modded out by their maximal compact subgroups $H$. The Lie algebra
$\mathfrak{g}$ of the group $G$ is a semi-direct sum:
\begin{equation}
\mathfrak{g} =   \mathfrak{g}^0\oplus \mathfrak{g}^{+1}
\end{equation}
\begin{eqnarray}
 \mathfrak{g}^0 &=& \mathfrak{so}_{\lambda}(1,1)  \oplus \mathfrak{so}(q+1,1) \oplus {\cal S}_q(P,Q) \nonumber\\
\mathfrak{g}^{+1} &= &(s,v)
\end{eqnarray}
Here, ${\cal S}_q(P,Q)$ is a Lie algebra generated by the metric
preserving elements of the centralizer of the Clifford algebra
representation, whereas $s$ denotes a spinor representation of
$so(q+1,1)$ (of dimension ${\cal D}_{q+1}$), and $v$ denotes the
vector representation of ${\cal S}_q(P,Q)$, which is of dimension
$(P+Q)$.\footnote{We
 should note that in case the scalar manifold  is a symmetric
space the above Lie algebra gets extended by additional symmetry generators belonging to grade $-1$ space
transforming in the conjugate representation of
 $ g_{+1}$ with respect to $g_{0}$.
 The generic non-Jordan  family corresponds to theories with $q=-1$ and ${\cal S}(P)= \mathfrak{so}(P)$. On the other hand,
 for the generic Jordan family on has $q=0$ and ${\cal S}(P)= \mathfrak{so}(P)$. The magical theories correspond to
 $q=1,2,4,8$ and $P=1$ \cite{dwvp92}. }
 The isotropy group $H$ is
\begin{equation}
H=SO(q+1)\otimes {\cal S}_q(P,Q)
\end{equation}
The  possible groups ${\cal S}_q(P,Q)$ and the associated real Clifford algebras were given in \cite{dWvP1} which
we list in Table 5.

\begin{table}[htb]
\begin{center}
\begin{tabular}{||c|c|c|l||}\hline
$q$  &${\cal C}(q+1,0)$& ${\cal D}_{q+1}$&${\cal S}_q(P,Q)$
\\ \hline
$-1$ &$\Rbar$    &1         &$SO(P)$     \\
0    &$\Rbar\oplus \Rbar $&1&$SO(P)\otimes SO(Q)$ \\
1    &$\Rbar(2)$ &2         &$SO(P)$     \\
2    &$\Cbar(2)$ &4         &$U(P)$     \\
3    &$\Hbar(2)$ &8         &$USp(2P)$\\
4    &$\Hbar (2)\oplus \Hbar (2)$&8&$USp(2P)\otimes USp(2Q)$\\
5    &$\Hbar(4)$ &16&$USp(2P)$   \\
6    &$\Cbar(8)$ &16&$U(P)$     \\
7    &$\Rbar(16)$&16&$SO(P)$     \\
$n+8$ &  $\Rbar(16)\otimes{\cal C}(n+1,0)$&16 ${\cal D}_n$ &
as for $q=n$\\
\hline
\end{tabular}
\end{center}
\caption{Real Clifford algebras
 ${\cal C}(q+1,0)$ in $q+1$  Euclidean dimensions.
$\mathbb{A}(n) $ denotes the $n\times n$ matrices over the division
algebra $\mathbb{A}$ , while
 ${\cal D}_{q+1}$ denotes the
real dimension of an irreducible representation of the Clifford algebra. The ${\cal S}_q (P,Q)$ is the metric
preserving group in the centralizer of the Clifford algebra
 in the $(P+Q){\cal D}_{q+1}$  dimensional representation.}
\end{table}

Under dimensional reduction to four dimensions  the isometry groups
of the resulting homogeneous scalar manifolds extend to  duality
symmetry groups of the corresponding MESGTs. These isometry groups
are not semisimple for homogeneous and non-symmetric scalar
manifolds. Their Lie algebras $\mathfrak{W}$  have a graded
decomposition of the form \footnote{ Again for theories whose scalar
manifolds are symmetric spaces in four dimensions the full Lie
algebras of isometries have , in addition, generators of grade $-1$
and $-2$, corresponding to hidden symmetries.}
\begin{equation}
\mathfrak{W}= \mathfrak{W}^0 \oplus \mathfrak{W}^1 \oplus
\mathfrak{W}^2
\end{equation}
 with respect to the $SO(1,1)_{\alpha}$ whose generator
 is
 \[
 \alpha= \frac{2}{3} \lambda + \beta
 \]
 where $\lambda$ is the generator of $SO(1,1)_{\lambda}$ that
 determines the graded decomposition in five dimensions given above.
For the 4D theories that come from the 5D theories listed in Table 5
the corresponding isometries are
\[
\mathfrak{W}^0 = \mathfrak{so}(1,1)_{\alpha} \oplus
\mathfrak{so}(q+2,2) \oplus {\cal S}_q (P,Q)
\]
\[
\mathfrak{W}^1 =(1,S,V)
\]
\[ \mathfrak{W}^2 = (2,0,0)
\]
where $S$ is the spinor representation of $SO(q+2,2)$ and $V$
represents the ($P+Q$) dimensional vector representation of ${\cal
S}_q (P,Q)$. The grade $+2$  subspace is again one dimensional.
 Denoting the group generated by $\mathcal{W}$ as $G$ one finds that
 the 4D scalar manifolds are the coset spaces
 \[ \mathcal{M}_4 = \frac{G}{H} \]
 where $H = SO(q+2)\times SO(2) \times {\cal S}_q (P,Q) $ \footnote{
 Following \cite{dwvp92} we label the group generated by ${\cal S}_q
 (P,Q)$ with the same symbol.}.

 The maximal
semisimple subgroups of the full isometry groups for homogeneous and
nonsymmetric scalar manifolds in four dimensions is then of the form
\[ SO(q+2,2) \times {\cal S}_0(P,Q) \]
where ${\cal S}_0(P,Q) $ denotes the semisimple subgroup of ${\cal
S}(P,Q)$ listed in Table 4.

In four dimensions we have $ \tn +2= 4 + q + (P+Q) {\cal D}_{q+1} $
vector fields. Under the action of $SO(q+2,2) \times {\cal S}(P,Q)$
the field strengths of these vector fields and their magnetic duals
transform as follows

\[ ( \mathcal{F}^{A\;\mu\nu} \oplus \mathcal{G}_{A\;\mu\nu} ) \cong
 (q+4,1) + (q+4,1) + (S,P+Q)
\]

where $S=2{\cal D}_{q+1}$.  Now $q\geq -1$ and for homogeneous and
nonsymmetric scalar manifolds $P\geq 1 $ and $Q \geq 1$
\cite{dwvp92}. Hence in the corresponding theories the isometry
groups do not have  any simple subgroups under which all the vector
field strengths transform irreducibly. Therefore they are not
unified MESGTs.

\subsection{Theories defined by Lorentzian Jordan algebras}

Let us now consider the 4D  MESGTs defined by Lorentzian Jordan
algebras.  The manifest isometries of the 4D scalar manifolds,
$\mathcal{M}_4$, of these theories   generate the semidirect product
groups  of the form
\begin{equation}
[\textrm{Aut}(J_{(1,N)}^{\mathbb{A}})\times SO(1,1)] \circledS
T_{J_{(1,N)_0}^{\mathbb{A}}}
\end{equation}
where $T_{J_{(1,N)_0}^{\mathbb{A}}}$ denotes the translations by
traceless elements of $ J_{(1,N)}^{\mathbb{A}}$. These isometries
also form  the invariance groups of the Lagrangians of
 the corresponding MESGTs (modulo shifts in the theta angles).
The Lie algebras of these symmetry groups have the graded
decomposition
\[ \mathfrak{g} = \mathfrak{g}^0 \oplus \mathfrak{g}^1, \]
where the grading is with respect to the generator of scale
transformations $SO(1,1)$, as discussed in the previous section, and
$\mathfrak{g}^0$ is the Lie algebra of
$\textrm{Aut}(J_{(1,N)}^{\mathbb{A}})\times SO(1,1)] $.
 An important question is whether or not there are further ``hidden''
isometries, in particular with negative scale dimension with respect
to $SO(1,1)$. For   $\mathcal{N}=2$   MESGTs defined by Lorentzian
Jordan algebras $ J_{(1,3)}^{\mathbb{A}} $ of degree 4, where
$\mathbb{A}= \mathbb{R},\mathbb{C}$ and $\mathbb{H}$ the cubic forms
defined by the structure constants of the traceless elements
coincide  with the norm forms of Euclidean Jordan algebras of degree
3 over $\mathbb{C}, \mathbb{H}$ and $\mathbb{O}$, respectively. The
isometry groups of the scalar manifolds of theories defined by
Euclidean Jordan algebras of degree 3 are simply the conformal
groups of the corresponding Euclidean Jordan algebras. Therefore in
these three theories the Lie algebra of the isometry group has a
3-graded decomposition

\[ \mathfrak{isom}(\mathcal{M}_4) = \mathfrak{K}_{J_3^{\mathbb{A}}} \oplus  \mathfrak{str}(J_3^{\mathbb{A}})
\oplus \mathfrak{T}_{J_3^{\mathbb{A}}} \]

where $\mathfrak{K}$  represent special conformal generators of
negative dimension and $\mathfrak{T}$ denote translations. Since the
cubic forms defined by the structure constants of the traceless
elements of the Lorentzian algebras $ J_{(1,3)}^{\mathbb{A}} $ are
not their generic norm forms, which are quartic, their  conformal
symmetry algebras are not the symmetries of the corresponding four
dimensional MESGTs. To highlight the differences in the symmetry
algebras of the Lorentzian Jordan algebras $ J_{(1,3)}^{\mathbb{A}}
$ and those of the corresponding Euclidean Jordan algebras of degree
three we list them in Table 6.

\begin{table}[ht]
\begin{center}
\begin{displaymath}
\begin{array}{|c|c|c|c|}
\hline
~&~&~&~\\
J &\textrm{Aut}(J) & \textrm{Str}_0(J) & \textrm{Conf}(J)   \\
\hline
~&~&~&~\\
J_{(1,3)}^{\mathbb R}& SO(3,1) & SL(4,\mathbb{R}) & Sp(8,\mathbb{R})
\\
~&~&~&~\\
J_3^{\mathbb{C}} & SU(3) & SL(3,\mathbb{C}) & SU(3,3) \\ \hline
~&~&~&~ \\
J_{(1,3)}^{\mathbb C} & SU(3,1) & SL(4,\mathbb{C})& SU(4,4) \\
~&~&~&~\\
J_3^{\mathbb{H}} & USp(6) & SU^*(6) & SO^*(12) \\ \hline
~&~&~&~ \\
J_{(1,3)}^{\mathbb H}& USp(6,2) & SU^*(8)  & SO^*(16)    \\
~&~&~&~\\
J_3^{\mathbb{O}} & F_4 & E_{6(-26)} & E_{7(-25)}  \\
 \hline
\end{array}
\end{displaymath}
\end{center}

\caption{List of the simple Lorentzian Jordan algebras of type
$J_{(1,3)}^{\mathbb{R}},J_{(1,3)}^{\mathbb{C}} ,
J_{(1,3)}^{\mathbb{H}}$ and the corresponding Euclidean Jordan
algebras $J_3^{\mathbb{C}} , J_3^{\mathbb{H}} ,J_3^{\mathbb{O}} $
and their automorphism, reduced structure and conformal (M\"{o}bius)
groups.}
\end{table}

 For magical supergravity
theories the global symmetry group of the five-dimensional theory is
the reduced structure group $\textrm{Str}_0(J)$ of the underlying Jordan
algebra $J$ and the number of vector fields $\tn$ is equal to the
dimension        $\textrm{dim}(J)$ of $J$. Under dimensional reduction to 4D
there is a vector field that is a singlet of $\textrm{Str}_0(J)$, namely the
vector field coming from the graviton in five dimensions. For these
theories the full U-duality group is the conformal group of $J$
under whose action electric and magnetic fields strengths transform
in the irreducible $(2\tn +2)$ dimensional representation.

Viewing three of the  magical theories as coming from the Lorentzian
Jordan algebras $J_{(1,3)}^{\mathbb{A}}$ ($\mathbb{A}= \mathbb{R},
\mathbb{C}, \mathbb{H}), $ the manifest symmetry groups of the   5D
theories are the automorphism groups of $ J_{(1,3)}^{\mathbb{A}}$.
However, they have   ``hidden'' symmetries in five dimensions that
extend the automorphism groups $ J_{(1,3)}^{\mathbb{A}}$
($\mathbb{A}= \mathbb{R}, \mathbb{C}, \mathbb{H}) $  to the  reduced
structure groups $\textrm{Str}_0(J_3^{\mathbb{C}})$ ,
$\textrm{Str}_0(J_3^{\mathbb{H}}) $ and
$\textrm{Str}_0(J_3^{\mathbb{O}} )$, respectively. By contrast, the
reduced structure groups of the Lorentzian Jordan algebras $
J_{(1,3)}^{\mathbb{A}}$ are not symmetry groups of the corresponding
five-dimensional MESGTs.

  The number of vector fields of the   four-dimensional  MESGTs defined by
   Lorentzian Jordan algebras
  $ J_{(1,3)}^{\mathbb{A}}$ is
equal to the dimension of $ J_{(1,3)}^{\mathbb{A}}$. Under the
action of their reduced structure groups the elements of $
J_{(1,3)}^{\mathbb{A}}$  transform in real irreducible
representations.  As discussed above, there exist symplectic
sections of the corresponding four-dimensional MESGTs such that the
reduced structure groups of $J_{(1,3)}^{\mathbb{A}}$ become symmetries
of their Lagrangians. This is why  they are unified MESGTs in four
dimensions as well. The important fact to note is that the reduced
structure groups of the Lorentzian Jordan algebras $
J_{(1,3)}^{\mathbb{A}} $ are the maximal common subgroups of the
conformal groups of $ J_{(1,3)}^{\mathbb{A}} $ ($
\mathbb{A}=\mathbb{R}, \mathbb{C}, \mathbb{H}) $   and of the
conformal groups of the corresponding Euclidean Jordan algebras
$J_3^{\mathbb{A'}}$ ($\mathbb{A'}=\mathbb{C},\mathbb{H},
\mathbb{O}$), respectively. The latter groups are the full U-duality
groups of these theories in four dimensions.

For the five-dimensional  $\mathcal{N}=2$
 MESGTs defined by the Lorentzian Jordan
algebras $J_{(1,N)}^{\mathbb{A}}$ for $N \neq 3$, the scalar
manifolds are not  homogeneous. Hence the analysis of \cite{dwvp92}
regarding the existence of negative grade isometry generators does
not apply. However, one can  prove that, also for these theories,
there cannot  be any negative dimension isometry generators of the
type that comes up in symmetric space  theories as follows.

 From the discussion in the
previous section about the geometry of these theories  it follows
that the Lie algebra $[\mathfrak{der}(J_{(1,N)}^{\mathbb{A}}) \oplus
\mathfrak{\beta}] \,\circledS\, \mathfrak{T}_{J_{(1,N)_0}}$ of the
manifest symmetry group $[\textrm{Aut}(J_{(1,N)}^{\mathbb{A}})\times
SO(1,1)]\,\circledS\,T_{J_{(1,N)_0}^{\mathbb{A}}}$ can be embedded
in the natural 3-graded decomposition of the conformal algebra of
$J_{(1,N)}^{\mathbb{A}}$ as follows:
\begin{equation}
\mathfrak{conf}(J_{(1,N)}^{\mathbb{A}}) = [ \mathfrak{K}_I \oplus
\mathfrak{K}_{J_{(1,N)_0}}] \oplus [\mathfrak{der}
(J_{(1,N)}^{\mathbb{A}}) \oplus \beta \oplus
\mathfrak{M}(J_{(1,N)_0}^{\mathbb{A}})] \oplus [
\mathfrak{T}_{J_{(1,N)_0}}^{\mathbb{A}} + \mathfrak{T}_I ]
\end{equation}
where $I$ denotes the identity element of the Jordan algebra and
$\mathfrak{M}(J_0) $ denotes multiplications by traceless elements
of $J$. The conformal Lie algebras of simple Jordan algebras of
dimension greater than two are simple. From the commutation
relations of the generators of conformal algebras of simple Jordan
algebras \cite{mg92},  it follows that any single special conformal
generator $\mathfrak{K}_a$ of negative dimension will extend the
manifest symmetry algebra $[\mathfrak{der}(J_{(1,N)}^{\mathbb{A}})
\oplus \mathfrak{\beta}] \circledS \mathfrak{T}_{J_{(1,N)_0}}$ to
the full conformal algebra $\mathfrak{conf}(J_{(1,N)}^{\mathbb{A}})$
 under
commutation. Now the number of vector fields in 4D theories defined
by Lorentzian Jordan algebras is equal to the dimension of the
Jordan algebra itself:
\[ l \equiv \dim (J_{(1,N)}^{\mathbb{A}}) = N+1+ \frac{1}{2}N(N+1) \dim (\mathbb{A}) \]
However, the conformal group of $J_{(1,N)}^{\mathbb{A}}$  acts
nonlinearly on $J_{(1,N)}^{\mathbb{A}}$ and does not have any
irreducible {\it linear} representation of dimension $l$, which
decomposes into a singlet plus an irrep of dimension $(l-1)$ under
its automorphism group. If the conformal group were to act as the
full on-shell U-duality   that transforms field strengths and their
magnetic duals into each other, it would have to have a linear
representation of dimension $2l$. This linear representation of
dimension $2l$ must then decompose  as two singlets and two irreps
of dimension $(l-1)$ under the automorphism group of the Jordan
algebra. Going through the list of the conformal groups of simple
Lorentzian Jordan algebras it is straightforward to show that they
do not admit such a linear representation. Thus, for 4D,
$\mathcal{N}=2$ MESGTs defined by the Lorentzian Jordan algebras
$J_{(1,N)}^{\mathbb{A}}$  their  conformal groups can not be
symmetries   on-shell or off-shell. Hence, there cannot be any
symmetry generators of negative dimension corresponding to special
conformal transformations of the corresponding Jordan algebras.

Even though there cannot  be any negative dimension hidden symmetry
generators associated with conformal transformations in theories
defined by Lorentzian Jordan algebras $J_{(1,N)}^{\mathbb{A}}$  (for
$N\neq3$) in four dimensions, one may wonder if there could be extra
hidden symmetry generators of dimension zero. In particular if extra
symmetry generators corresponding to multiplications by traceless
elements of the Jordan algebra were to arise as hidden symmetries
under dimensional reduction they would extend the five-dimensional
manifest symmetry group $\textrm{Aut}(J_{(1,N)})$ to the reduced
structure group that acts irreducibly on the Jordan algebra whose
dimension $l$ is equal to the number of vector fields in the theory.
However the extension of the grade zero symmetry algebra to that of
the reduced structure group would require that the number of
translation generators  be increased from $l-1$ to $l$ (see
appendix A). But we have  only $l-1$ additional scalars that
come from the vector fields in 5d. Therefore we can rule out the
extension of the automorphism group to the reduced structure group
within the symplectic section one obtains by dimensional reduction.
 However, in general the symplectic sections that
make manifest a unifying symmetry group of the Lagrangian break the
translational symmetry. For example, in the exceptional supergravity
theory the unifying symmetry group is $SU^*(8)$ under which the
remaining generators of the full duality group $E_{7(-25)}$
transform in the 70 dimensional representation, whose action does
not leave the Lagrangian invariant!  Therefore, for the
$\mathcal{N}=2$ MESGTs defined by Lorentzian Jordan algebras
$J_{(1,N)}^{\mathbb{A}}$  ($N\neq 3$),  we can not rule out the
existence of symplectic sections where their reduced structure
groups become unifying symmetry groups of their Lagrangians. What
makes the analysis of these theories more complicated is the fact
that their scalar manifolds are not homogeneous. We leave this as an
open problem for future investigations.

 We finally note that,  under dimensional reduction to  4D,  one gets an extra vector
multiplet from the 5 dimensional graviton supermultiplet.  One may
ask if it possible to obtain unified MESGTs in 4D by first
dimensionally reducing the 5d unified MESGTs  to four dimensions
followed by a  truncation of this extra vector supermultiplet. As we
discussed at the beginning of this section the scalars of this extra
vector supermultiplet parameterize the manifold \[
\frac{SU(1,1)_G}{U(1)}. \] For the theories defined by Euclidean
Jordan algebras we listed the centralizers of the group $SU(1,1)_G$
in section 5.2.
 After truncation  the centralizing groups would become the U-duality
groups. It is manifest from the list of centralizing groups that,
even if the truncations are consistent, the resulting theories can
not be unified since there will always be a singlet vector field of
the resulting U-duality groups. The same analysis can be extended to
the theories defined by Lorentzian Jordan algebras of degree greater
than four. In this case the $SU(1,1)_G$ is not part of the full
U-duality group in four dimensions, but rather its parabolic
subgroup $SO(1,1)\circledS T$ is. It can be shown that the
centralizer of this parabolic subgroup will have a singlet vector
field after the truncation, even if the truncation is consistent.

\end{itemize}
\section{Conclusion  and summary}

In this paper, we studied  the    four-dimensional  unified
$\mathcal{N}=2$ MESGTs that can be obtained by dimensional reduction
from 5D and their possible  gaugings so as to obtain unified
YMESGTs. We found that the generic Jordan family of MESGTs that are
not unified in 5D become unified in 4D. The electric subgroup
of the full duality group is $SO(\tn,2)$   under which all the
vector field strengths, including that of the graviphoton, transform
irreducibly. Of these theories only one of them (corresponding to
$\tn=1$) can be gauged so as to obtain a unified YMESGT with the
gauge group $SO(2,1)=SU(1,1)$. This gauging breaks the full duality
group $SO(2,1)\times SO(2,1)$ down to $SO(2,1)$. The four magical
supergravity theories defined by simple Jordan algebras of degree
three are unified MESGTs in 5D. In 4D they all have simple
duality groups. However only the theories defined by
$J_3^{\mathbb{C}}$ , $J_3^{\mathbb{H}}$ and $J_3^{\mathbb{O}}$ are
unified MESGTs in 4D , with unifying symmetry groups
$SL(4,\mathbb{R}), SL(4,\mathbb{C})$ and $SU^*(8)$, respectively.
The $Sp(4,\mathbb{R})=SO(3,2)$ subgroup of the full duality group
$SU(3,3)$ of the $J_3^{\mathbb{C}}$ theory can ge gauged so as to
obtain a unified YMESGT in 4D. The theory defined by
$J_3^{\mathbb{H}}$ can not be gauged so as to obtain a unified
YMESGT. In the exceptional supergravity defined by the exceptional
Jordan algebra $J_3^{\mathbb{O}}$ one can gauge the
$SO^*(8)=SO(6,2)$ subgroup of the duality group $E_{7(-25)}$ to get
a unified YMESGT in 4D. These results are summarized in Table 4.

The generic non-Jordan family and the theories whose scalar
manifolds are homogeneous but not symmetric do not lead to unified
MESGTs in 4D.

The unified five-dimensional MESGTs defined by Lorentzian Jordan
algebras, $J_{(1,N)}^{\mathbb{A}}$, of degree $p=(N+1)$, ($N\neq3$)
do not have hidden symmetries associated with special conformal
transformations of   $J_{(1,N)}^{\mathbb{A}}$. Whether they admit
other hidden symmetries that admit symplectic sections with their
reduced structure groups acting as their unifying symmetry groups is
an open problem.

 In this paper, we considered  the 4D, $\mathcal{N}=2$
MESGTs that originate from five dimensions. We should perhaps
mention that there is a family of 4D MESGTs whose scalar manifolds
are the symmetric spaces \cite{luciani} \[ \frac{SU(n,1)}{U(n)}
\] which do not originate from five dimensions. They are unified MESGTs since all
the vector field strengths transform irreducibly under the electric
subgroup $SO(n,1)$. Of these only the theory with $n=2$ can be
gauged so as to obtain a unified YMESGT with the gauge group
$SO(2,1)$. This unified YMESGT is different from the $SO(2,1)$
gauged unified YMESGT considered in section 5, since their scalar
manifolds are different, namely   one has
\[ SU(2,1)/U(2) \]  and
\[ \frac{SU(1,1)\times SO(2,1)}{U(1)\times U(1)}. \]

\section*{Appendix}
\appendix
\section{ Symmetry algebras of Jordan algebras}

 Jordan  algebras are   commutative  \[ a \circ b = b \circ a \] and, in general,  non-associative
 algebras that satisfy   the identity
\[
(a\circ b)\circ a^2 = a\circ(b\circ a^2) \,. 
\]
Elements of a  Jordan algebra  form a so-called Jordan triple system
(JTS) under the Jordan triple product
\[
 \JTP{a}{b}{c} =  a\circ (b\circ c) +(a\circ b)\circ c
 - b \circ(a\circ c) \, . \nn
\]
 This triple product satisfies the identities
\[
\begin{array}{l}
 \JTP{a}{b}{c} = \JTP{c}{b}{a}  \,, \\[1ex]
 \JTP{a}{b}{\JTP{c}{d}{x}}-\JTP{c}{d}{\JTP{a}{b}{x}}
-\JTP{a}{\JTP{d}{c}{b}}{x}+\JTP{\JTP{c}{d}{a}}{b}{x} = 0 \,       ,
\end{array}
\]
 which are  the defining identities of a Jordan triple system.

The Lie algebra $\mathfrak{aut}(J)$ of the automorphism group of a
Jordan algebra $J$ is simply its derivation algebra
$\textrm{Der}(J)$.
Derivations are defined as linear transformations $D$ that satisfy
the Leibniz rule; i.e., given $a,b \in J$
\[ D(a\circ b) = (Da)\circ b + a \circ (Db) \in J \]
They close and form a Lie algebra under commutator product.
Derivations of a Jordan algebra can always be written in the form
\cite{mccrimmon}
\[ D_{a,b} \equiv [ M_a,M_b]  ;  a,b \in J \]
where $M_a$ denotes multiplication by element $a$: \[ M_a x = a\circ
x \] Clearly       the multiplications  $M_J$ close into derivations under commutation. The  Lie algebra $\mathfrak{str}(J)$ of the
structure group of $J$  is then generated by derivations and multiplications
$\textrm{M}(J)$ by the elements of $J$
\[ \mathfrak{str}(J) = \textrm{Der}(J) \oplus \textrm{M}(J). \]
Multiplications by \emph{traceless} elements of $J$ together with
derivations generate the reduced structure group $\textrm{Str}_0(J)$ , which
turns out to be the invariance group of  norm of the  Jordan algebra
$J$.

 For a Euclidean Jordan algebra derivations generate a compact
automorphism group and the multiplications $M(J)$ by the elements of
$J$ correspond to  non-compact generators of $\mathfrak{str}(J)$. For
a noncompact, e.g., Lorentzian,  Jordan algebra the automorphism group
is noncompact and hence some of the derivations correspond to
noncompact generators. Similarly multiplications $M_a$ by an element
$a \in J$ will generate compact or noncompact transformation
depending on whether the element $a$ is anti-Hermitian or hermitian,
respectively. Even though the automorphism groups of Euclidean
Jordan algebras $J_{(N+1)}^{\mathbb{A}}$ and Lorentzian Jordan
algebras $J_{(1,N)}^{\mathbb{A}}$ are not isomorphic, their
structure algebras are:

\[
\mathfrak{str}(J_{(N+1)}^{\mathbb{A}}) \thickapprox
\mathfrak{str}(J_{(1,N)}^{\mathbb{A}}) .
\]

 Koecher \cite{koecher} introduced  the concept of linear fractional
transformation groups  of Jordan algebras. The linear fractional
groups can be interpreted as generalized conformal groups of
spacetimes coordinatized by Jordan algebras \cite{MG75}, which we
denote as $\textrm{Conf}(J)$. Their Lie algebras can be given a 3-graded
decomposition \cite{tkk}
\[ \mathfrak{conf}(J)= g^{-1} \oplus g^{0} \oplus g^{+1} = K_J \oplus \mathfrak{str}(J)
\oplus T_J, \] where the grade zero subalgebra is simply the
structure algebra and grade +1 and -1 subspaces are spanned by
translations $T_J$ and special conformal generators $K_J$,
respectively. Within this framework the reduced structure group is
simply the generalized Lorentz group $\textrm{Lor}(J)$ of the
spacetime defined by  $J$ \cite{MG75}, and, in addition, the structure group
includes the dilatations.

 The Tits-Kantor-Koecher (TKK)
construction~\cite{tkk} associates with every JTS  a 3-graded Lie
algebra \footnote{With the exception of the Lie algebras of $G_2 ,
F_4 $ and $E_8$ every simple Lie algebra $\mathfrak{g}$ can be given
a three graded decomposition with respect to a subalgebra
$\mathfrak{g}^0$ of maximal rank.}
\[
  \fg = \fg^{-1} \oplus \fg^{0} \oplus\fg^{+1} \,, \label{3-grading}
\]

In the TKK construction the elements  of the $\fg^{+1}$ subspace of
the Lie algebra   $\fg$  are labeled by the elements $a\!\in\!J$:

\[ U_a \in \fg^{+1} \Leftrightarrow a \in J \]

Furthermore every such Lie algebra \fg ~ admits an involution
$\tilde{}$ , which maps the elements of the grade $+1$ space onto
the elements of the subspace of grade $-1$:
\[
  \tilde{}  :  U_a \in  \fg^{+1} \Rightarrow \tilde{U}_a \in \fg^{-1}
\]
To get the  complete set of generators of \fg ~ one  defines further
\[
[U_a,\tilde{U}_b] =: S_{ab}  \]  \[ [S_{ab},U_c] =: U_{\{abc\}}
\]
where $S_{ab}\in\fg^0$ and $\{abc\}$ is the Jordan triple product
under which  $J$ is closed.

The remaining commutation relations are
\[
[S_{ab},\tilde{U}_c]    = \tilde{U}_{\{bac\}} \]

\[ [S_{ab},S_{cd}]
= S_{\{abc\}d} - S_{c\{bad\}},
\]
and the Jacobi identities follow from the defining identities of a
JTS given above.

The Lie algebra generated by $S_{ab}$ is simply  the structure
algebra of  $J$ since
\[ S_{ab} = D_{a,b} + M_{a\circ b} \]
where $D_{a,b} = [M_a,M_b] $ is a derivation and $M_{a\circ b}$ is
multiplication by the element $a\circ b$. The traceless elements of
this action of $S_{ab}$ generate the reduced structure algebra of
$J$.

The above  concepts are best illustrated in terms of a simple and
familiar example, namely the conformal group in four dimensions, and
its realization via the Jordan algebra $J_2^{\mathbb{C}}$ of
hermitean $2 \times 2$ matrices over $\mathbb{C}$. As is well known,
these matrices are in one-to-one correspondence with four-vectors
$x^\mu$ in Minkowski space \[ x \equiv x_\mu \sigma^\mu \in
J_2^{\mathbb{C}} \]  where $\sigma^\mu := (1, \vec{\sigma})$. The
``norm form'' on this Jordan algebra is just the ordinary
determinant, i.e.
\[
N(x) := \det (x) = x_\mu x^\mu
\]

The automorphism group of $J_2^{\mathbb{C}}$ is just the rotation
group $SU(2)$, its   structure group is the product
$SL(2,\mathbb{C}) \times \mathcal{D}$, i.e. the Lorentz group times
dilatations. The conformal group of $J_2^{\mathbb{C}}$ is $su(2,2)$,
which has  a three-graded structure
\[ \mathfrak{g} = \mathfrak{g}^{-1} \oplus \mathfrak{g}^{0}
                   \oplus\mathfrak{g}^{+1}
\]
where the grade $+1$ and grade $-1$ spaces correspond to generators
of translations $P^\mu$ and special conformal transformations
$K^\mu$, respectively, while the grade 0 subspace is spanned by the
Lorentz generators $M^{\mu\nu}$ and the dilatation generator $D$.
The subspaces $\fg^{+1}$ and $\fg^{-1}$ can both be labelled by the
elements of the Jordan algebra $J_2^{\mathbb{C}}$.  Letting  $a =
a_\mu \sigma^\mu$ and $ b= b_\mu \sigma^\mu$  we can expand the
grade +1 and -1 subspaces in terms of  the generators of
translations $P_\mu$ and special conformal transformations $K_\mu$
as
\[
U_a := a_\mu P^\mu \in \fg^{+1} \] \[ \tilde U_b := b_\mu K^\mu \in
\fg^{-1} \,.
\]
The generators in $\mathfrak{g}^0$ are labeled as
\[
S_{a b} := a_\mu b_\nu ( M^{\mu\nu} + \eta^{\mu\nu} D) \,.
\]
The conformal group is realized non-linearly on the space of
four-vectors $x\in J_2^{\mathbb{C}}$, with a M\"obius-like action,
which  acquires a very simple form when expressed in terms of the
Jordan triple product, namely
\[
U_a (x) = a \]
\[ S_{ab} (x) =  \JTP{a}{b}{x}\] \[ \tilde U_c
(x) = -\frac{1}{2} \JTP{x}{c}{x} \,,
\]
Such a realization extends to all JTS's defined by Jordan algebras
\cite{MG75}.

\section{Symplectic formulation
of $N=2$ MESGTs in four dimensions}
The          self-dual  field strengths $F_{\mu\nu}^{A +}$ and their
magnetic
 ``duals''  \cite{duality,fre},
\begin{equation}
G_{\mu\nu A}^{+}:=\frac{\delta \mathcal{L}^{(4)}}{\delta F_{\mu\nu}^{A+}}=-\frac{i}{2}\mathcal{N}_{AB}F^{\mu\nu B + },
\end{equation}
can be combined into a symplectic vector
\begin{equation}
 \left( \begin{array}{c}
 F_{\mu\nu}^{A +} \\
 G_{\mu\nu B}^{ + }
 \end{array}  \right)
\end{equation}
so that the equations of motion that follow from (\ref{redlag1b})
are invariant under the
(global)      symplectic rotations
\begin{equation}
 \left( \begin{array}{c}
 X^{A} \\
 F_{B}
 \end{array}  \right)  \longrightarrow  \mathcal{O} \left( \begin{array}{c}
 X^{A} \\
 F_{B}
 \end{array}  \right)  , \qquad   \left( \begin{array}{c}
 F_{\mu\nu}^{A +} \\
 G_{\mu\nu B}^{ + }
 \end{array}  \right)   \longrightarrow  \mathcal{O}
   \left( \begin{array}{c}
 F_{\mu\nu}^{A +} \\
 G_{\mu\nu B}^{ + }
 \end{array}  \right)
\end{equation}
with $\mathcal{O}$ being  a symplectic matrix with respect to the
symplectic metric
\begin{equation}
\omega =  \left( \begin{array}{cc}
0 & \delta_{B}^{A} \\
- \delta_{B}^{A} & 0
\end{array}  \right)  .
\end{equation}
Writing $\mathcal{O}$ as
\begin{equation}
\mathcal{O} =   \left(  \begin{array}{cc}
A & B \\
C & D
  \end{array}      \right)  ,
\end{equation}
the period matrix $\mathcal{N}$ transforms as
\begin{equation}
\mathcal{N} \longrightarrow  (C + D \mathcal{N} ) ( A + B \mathcal{N} )^{-1}  .
\label{Ntrafo}
\end{equation}
Symplectic transformations with $B\neq 0$ correspond to electromagnetic duality transformation,  whereas transformations with $C\neq 0$ involve
shifts of the theta angles in the Lagrangian.

General symplectic tranformations will take a Lagrangian
$\mathcal{L}(F,G)$ with the field strengths
satisfying the Bianchi identities $dF^{A}=0$ and
$dG_{A}=0$ to a Lagrangian
$\tilde{\mathcal{L}}(\tilde{F},\tilde{G})$ with
the new field strengths satisfying $d{\tilde{F}}^{A}=0$ and
$d{\tilde{G}}_{A}=0.$  Theories related by such general
symplectic tranformations are called dual theories.

The subgroup, $\mathcal{U}$, of $Sp(2(\tn+2),\mathbb{R})$ that leaves the functional invariant \cite{dwvp92}
\[
\tilde{\mathcal{L}}(\tilde{F},\tilde{G})=\mathcal{L}(\tilde{F},\tilde{G}),
\]
is called  the duality invariance group (or ``U-duality group"). This is a symmetry group of the on-shell
theory (i.e., of the equations of motion).  A subgroup of the
duality invariance group that leaves the off-shell Lagrangian
invariant up to surface terms is called an ``electric subgroup",
$G_{E}$, since it transforms electric field strengths into electric field strengths only. Obviously, we have the inclusions \[ G_{E}\subset \mathcal{U}\subset
Sp(2(\tn+2),\mathbb{R}) ,
\]  and  the pure electric-magnetic transformations   are contained
in the coset $\mathcal{U}/G_{E}$, while dualizations leading to
different manifest electric subgroups  $G_{E}$ groups are contained
in $Sp(2n+4)/\mathcal{U}$ \cite{dWN}.

 For the symplectic formulation of
$\mathcal{N}=2$ MESGTs in 4D and further references we refer  the
reader to \cite{symplectic}.

{\bf Acknowledgement:}   M.Z. thanks Henning Samtleben   for discussions.   This work was supported in part by the
    National Science Foundation under grant number PHY-0245337. Any opinions, findings and conclusions or
     recommendations expressed in this material  are those of the author and do not necessarily
reflect the views of  the National Science Foundation. The work of M.Z. is supported    by the German Research Foundation (DFG) within the Emmy-Noether-Program
      (ZA 279/1-1).

\end{document}